\documentclass[fleqn,usenatbib]{aa}
\usepackage{txfonts}
\usepackage{graphicx}
\usepackage{natbib}
\usepackage{longtable}
\usepackage{subeqnarray}
\usepackage{cases}
\usepackage[colorlinks=true,citecolor=blue]{hyperref}
\usepackage{amsmath}
\usepackage[switch]{lineno} 
\let\oldequation\equation         
\let\oldendequation\endequation   
\renewenvironment{equation}{\linenomathNonumbers\oldequation}{\oldendequation\endlinenomath} 

\begin{document}
\title{Molecular Clouds at the Edge of the Galaxy}
\subtitle{I. Variation of CO\,$J$=2--1/1--0 Line Ratio}

\author{C. S. Luo\inst{\ref{inst1},\ref{inst2}}
\and X. D. Tang\inst{\ref{inst1},\ref{inst2},\ref{inst3},\ref{inst4}}
\and C. Henkel\inst{\ref{inst5},\ref{inst1}}
\and K. M. Menten\inst{\ref{inst5}}
\and Y. Sun\inst{\ref{inst6}}
\and Y. Gong\inst{\ref{inst6},\ref{inst5}}
\and X. W. Zheng\inst{\ref{inst7}}
\and D. L. Li\inst{\ref{inst1},\ref{inst2},\ref{inst4}}
\and Y. X. He\inst{\ref{inst1},\ref{inst2},\ref{inst4}}
\and X. Lu\inst{\ref{inst8}}
\and Y. P. Ao\inst{\ref{inst6}}
\and X. P. Chen\inst{\ref{inst6}}
\and T. Liu\inst{\ref{inst8}}
\and K. Wang\inst{\ref{inst9}}
\and J. W. Wu\inst{\ref{inst10},\ref{inst2}}
\and J. Esimbek\inst{\ref{inst1},\ref{inst3},\ref{inst4}}
\and J. J. Zhou\inst{\ref{inst1},\ref{inst3},\ref{inst4}}
\and J. J. Qiu\inst{\ref{inst11}}
\and X. Zhao\inst{\ref{inst1},\ref{inst2}}
\and J. S. Li\inst{\ref{inst1},\ref{inst2}}
\and Q. Zhao\inst{\ref{inst1},\ref{inst2}}
\and L. D. Liu\inst{\ref{inst1},\ref{inst2}}
}

\titlerunning{Variation of CO\,$J$=2--1/1--0 Line Ratio in the Galactic Edge Clouds}
\authorrunning{Luo et al.}

\institute{
Xinjiang Astronomical Observatory, Chinese Academy of Sciences, Urumqi 830011, PR China \label{inst1}\\
\email{tangxindi@xao.ac.cn}
\and University of the Chinese Academy of Sciences, Beijing100080, PR China \label{inst2}
\and Key Laboratory of Radio Astronomy and Technology, Chinese Academy of Sciences, 
A20 Datun Road, Chaoyang District, Beijing 100101, PR China \label{inst3}
\and Xinjiang Key Laboratory of Radio Astrophysics, Urumqi 830011, PR China \label{inst4}
\and Max-Planck-Institut f\"{u}r Radioastronomie, Auf dem H\"{u}gel 69, 53121 Bonn, Germany \label{inst5}
\and Purple Mountain Observatory, Chinese Academy of Sciences, Nanjing 210008, PR China \label{inst6}
\and School of Astronomy and Space Science, Nanjing University, Nanjing 210093, PR China \label{inst7}
\and Shanghai Astronomical Observatory, Chinese Academy of Sciences, 80 Nandan Road, Shanghai 200030, PR China \label{inst8}
\and Kavli Institute for Astronomy and Astrophysics, Peking University, Beijing 100871, PR China \label{inst9}
\and National Astronomical Observatories, Chinese Academy of Sciences, Beijing 100101, PR China \label{inst10}
\and School of Mathmatics and Physics, Jinggangshan University, Ji’an 343009, PR China \label{inst11}
}

\abstract{
The Galactic edge at Galactocentric distances of 14\,--\,22\,kpc provides an ideal laboratory to study molecular clouds in an environment that is different from the solar neighborhood,
due to its lower gas density, lower metallicity, and little or no perturbation from the spiral arms. Observations of CO\,($J$\,=\,2--1)
spectral lines were carried out towards 72 molecular clouds located at the Galactic edge using the IRAM\,30\,m telescope. 
Combined with CO\,($J$\,=\,1--0) data from the MWISP project, we investigate the variations of $R_{21}$ across these Galactic edge clouds, 
with $R_{21}$ representing CO(2-1)/CO(1-0) integrated intensity ratios. These are found to range from 0.3 to 3.0 with a mean of 1.0\,$\pm$\,0.1 in the Galactic edge clouds. 
The proportions of very low ratio gas (VLRG; $R_{21}$\,<\,0.4), low ratio gas (LRG; 0.4\,$\le$\,$R_{21}$\,<\,0.7), 
high ratio gas (HRG; 0.7\,$\le$\,$R_{21}$\,<\,1.0), and very high ratio gas 
(VHRG; $R_{21}$\,$\ge$\,1.0) are 6.9\%, 29.2\%, 26.4\%, and 37.5\%, respectively, indicating a significant presence of high $R_{21}$ ratio molecular gas within these regions. 
In our Galaxy, the $R_{21}$ ratio exhibits a gradient of initial radial decline followed by a high dispersion 
with increasing Galacticentric distance and a prevalence for high ratio gas. 
There is no apparent systematic variation within the Galactocentric distance range of 14 to 22\,kpc. 
A substantial proportion of HRG and VHRG is found to be associated with compact clouds and regions displaying star-forming activity, 
suggesting that the high $R_{21}$ ratios may stem from dense gas concentrations and recent episodes of star formation. 
}

\maketitle

\section{Introduction}
\label{sect:Introduction}
\subsection{Galactic edge clouds}
\label{sect:Galactic-Edge-Clouds}
The outer regions of the Galaxy offer a unique laboratory to investigate the influence of metallicity on star formation, 
with a spatial resolution that allows for the identification and classification of individual clumps and cores within giant molecular clouds (GMCs).
Compared to nearby clouds, little is known about molecular clouds at the edge of the Galaxy lying farther than $\sim$\,14\,kpc 
from the Galactic center. This is primarily due to the difficulty encountered in undertaking carbon monoxide (CO) surveys that are 
sensitive, unbiased, and cover a large area \citep{Digel1994}. 
The outer Galaxy may be younger than the regions farther inside \citep{Martig2016}.
The edge of the Galaxy appears to have a relatively shorter and 
less complex history of star formation, attributed to its lower gas density, lower metallicity, and minimal perturbation from the spiral arms
(e.g., \citealt{Smartt1997,Heyer2015,Wenger2019,Lian2023,Urquhart2024}). 
Consequently, this region provides an ideal laboratory to study the process of star and stellar cluster formation
without the complications arising from intertwined star formation activity in both space and time.

Various surveys of molecular lines (e.g., CO, CS, OH, HCO$^{+}$, HCN, H$_2$O, H$_2$CO, NH$_3$, CH$_3$OH)  in the outer Galaxy have been carried out 
(e.g., \citealt{Mead1988,Wouterloot1989,Wouterloot1993,Brand1994,Heyer1998,Pirogov1999,Snell2002,Ruffle2007,Blair2008,Dame2011,Sun2015a,Sun2018a,Wang2018,Braine2023,Koelemay2023}).
These surveys indicate that the molecular disk of our Galaxy extends 
beyond 20\,kpc from the Galactic center. The star formation process has been studied in the extreme outer Galaxy, which differs significantly from that of the solar neighborhood environment (e.g., \citealt{Digel1994,Brand1995,Brand2007,Wouterloot1996,Snell2002,Yasui2006, Yasui2008,Ruffle2007,Kobayashi2008,Izumi2014,Izumi2017,Izumi2022,Izumi2024,Armentrout2017,Matsuo2017,Sun2017,Sun2024a,Sun2024b,Shimonishi2021,Bernal2021,Braine2023,Urquhart2024,Lin2025}). 
However, the process of star formation in such environments has not been as extensively studied as in nearby star-forming regions 
due to their considerable heliocentric distances. 

Recently, \citet{Dame2011} and \citet{Sun2015a} identified a new segment of a spiral arm situated beyond the outer arm in both 
the first Galactic quadrant ($13^\circ<l<55^\circ$) and the second Galactic quadrant ($100^\circ<l<150^\circ$) (see Fig.\,\ref{fig:target}). 
A large number of molecular clouds have been identified on the new arm, which is located approximately 14\,--22\,kpc from the Galactic center \citep{Dame2011,Sun2015a}. 
These Galactic edge clouds present an optimal opportunity for investigating the properties and chemistry of molecular gas in the extreme outer Galaxy. 
Presently, the study of these molecular clouds still has been poor on basic properties, such as size, density, temperature, and star formation. 
A comprehensive investigation of the Galactic edge clouds will illuminate our comprehension of star formation within low-metallicity environments. 
Consequently, high-sensitivity CO observations with suitable angular resolutions are crucial for determining physical parameters. 
This will establish an observational basis for subsequent studies of rarer molecular tracers in the extreme outer Galaxy. 

\subsection{CO as a probe of physical conditions}
\label{sect:CO-Line-Ratio}
CO stands as the second most abundant molecular species, only following molecular hydrogen (H$_2$), and has consequently 
been employed as a reliable indicator of molecular gas. It is usually used to trace the variety of physical conditions of the 
bulk molecular gas (e.g., \citealt{Bolatto2013,Penaloza2017,Penaloza2018}). The two primary rotational transitions of the 
dominant CO molecule, $^{12}$C$^{16}$O\,$J$\,=\,1--0 (hereafter CO\,(1--0)) and $^{12}$C$^{16}$O\,$J$\,=\,2--1 (hereafter CO\,(2--1)), 
exhibit favorable frequencies for ground-based observations. These transitions are commonly employed to trace the mass of molecular gas in galaxies. 
Quantitative comparison of results obtained using the CO\,$J$=2--1/1--0 integrated intensity ratio, denoted as $R_{21}$, is gaining more and more importance. 
The $R_{21}$ ratio has been observed to exhibit variability in different environments. 
Investigations into this ratio have been conducted in a variety of molecular conditions, encompassing studies within the Milky Way 
(e.g., \citealt{Chiar1994,Sakamoto1994a,Sakamoto1997,Oka1998,Seta1998,Handa1999,Sawada2001,Salome2008,Yoda2010,Zhang2019}) 
and nearby galaxies 
(e.g., \citealt{Braine1992,Sorai2001,Bolatto2003,Leroy2009,Leroy2022,Koda2012,Koda2020,Zschaechner2018,denBrok2021,Yajima2021,Maeda2022,Liu2023,Keenan2025}). 
Through numerical simulations, it has been demonstrated that the $R_{21}$ effectively 
constrains the physical conditions of molecular gas, such as density, temperature, and optical depth 
(e.g., \citealt{Goldsmith1983,Castets1990,Sakamoto1994b,Sakamoto1997,Koda2012,Zhang2019,Schinnerer2024}).
Therefore, understanding the variations of $R_{21}$ perhaps in response to the ambient environment also holds the potential to yield valuable insights into the physical conditions of the molecular gas. 
Molecular gas can be categorized into four distinct groups based on the $R_{21}$ value using a large velocity gradient (LVG) approximation \citep{Sakamoto1997} in the following way: 
\begin{itemize}
\item
Very low ratio gas (VLRG) corresponds to cases where $R_{21}$\,<\,0.4. The faint VLRG is typically undetectable with observations of low sensitivity. 
Nevertheless, \citet{Krieg2008} identified a CO line ratio of $\sim$\,0.3, suggesting the presence of cold or sub-thermally excited gas within the overall galaxy host. 
\item
Low ratio gas (LRG) is observed for 0.4\,$\le$\,$R_{21}$\,<\,0.7. 
The excitation of the LRG has been demonstrated to be characterized by diffuse gas and/or gas of low kinetic temperatures. 
Measurements of $R_{21}$ indicate values of $\sim$\,0.5 for peripheral regions and $\sim$\,0.6 for intermediate regions 
in the Orion giant molecular cloud \citep{Sakamoto1994b}. The presence of the LRG is not confined solely to the disk and molecular 
ring of the Milky Way \citep{Chiar1994,Sawada2001}. It has also been detected in the interarm regions of galaxies
(e.g., \citealt{Koda2012,Koda2020}) and nearby star-forming galaxies (e.g., \citealt{denBrok2021,Yajima2021,Leroy2022,Maeda2022}). 
\item
High ratio gas (HRG) is defined as having a ratio of 0.7\,$\le$\,$R_{21}$\,<\,1.0. 
The excitation of the HRG is primarily due to collisions within local thermodynamic equilibrium (LTE), owing to the density and elevated temperature of the gas \citep{Sakamoto1994b}. 
This type of molecular characteristics has been found in the Large Magellanic Cloud (LMC) (e.g., \citealt{Sorai2001}), as well as in the central regions of normal star-forming galaxies (e.g., \citealt{Sawada2001,Leroy2009,Koda2012}) and the spiral arm regions of nearby galaxies (e.g., \citealt{Wiklind1990,Koda2012,Koda2020}). 
\item 
Very high ratio gas (VHRG) with values $R_{21}$\,$\ge$\,1.0:
The presence of VHRG is unexpected in instances of optically thick emission. 
Instead, it is typically found in either warm ($T_{\rm kin}\gtrsim50$\,K), dense ($n_{\rm H_2}\gtrsim 3\times10^{3}$\,cm$^{-3}$), and optically thin, or externally heated gas \citep{Sakamoto1997}. 
This gas has been detected in the vicinity of Orion\,KL \citep{Nishimura2015}, and within the N83/N84 molecular 
cloud complex situated inside the Small Magellanic Cloud (SMC) \citep{Bolatto2003}. 
Additionally, it has been identified in metal-poor galaxies such as NGC\,3310 \citep{Braine1992} and NGC\,1140 \citep{Hunt2017}. 
\end{itemize} 

As stated previously, $R_{21}$ exhibits consistent variations both within the Milky Way and across different galaxies, serving as an indicator of molecular gas conditions. 
Consequently, it is plausible that estimations of molecular gas mass, 
as well as related quantities and relationships derived from CO\,$J$=2--1 assuming a constant $R_{21}$\,$\sim$\,0.7 
(e.g., \citealt{Sandstrom2013,Sun2018c,Konig2021}), 
may be subject to misinterpretation. It is also crucial to acknowledge the existing uncertainty surrounding the CO-to-H$_2$ conversion factor.  
Therefore, it is advisable to conduct thorough examinations to validate the assumption of $R_{21}$ constancy and assess its 
impact on derived quantities and relationships pertaining to molecular gas. 

To date, a comprehensive measurement of the $R_{21}$ ratio in molecular clouds situated beyond 14\,kpc at the Milky Way's periphery remains unexplored. Hence, this investigation is particularly significant for numerous molecular clouds situated in the second Galactic quadrant at the edge of the Galaxy. We aim to explore the underlying causes of $R_{21}$ variations in conjunction with the physical properties of
molecular gas. In Sects.\,\ref{sect:Targets,observations,and data reduction} 
and \ref{sect:Results}, 
we introduce our targets, observations, data reduction, and describe the main results. The discussion of 
resulting $R_{21}$ ratios is presented in Sect.\,\ref{sect:discussion}. Our main conclusions are summarized in Sect.\,\ref{sect:summary}. 
This paper is part of the "Molecular Clouds at the Edge of the Galaxy" project, which focuses on the investigation of 
the physical and chemical properties of molecular clouds and star formation located in the outskirts of the Milky Way.

\section{Targets, observations, and data reduction}
\label{sect:Targets,observations,and data reduction}
\subsection{Targets}
\label{sect:Targets}
\begin{figure}[t]
\centering
\includegraphics[width=0.48\textwidth]{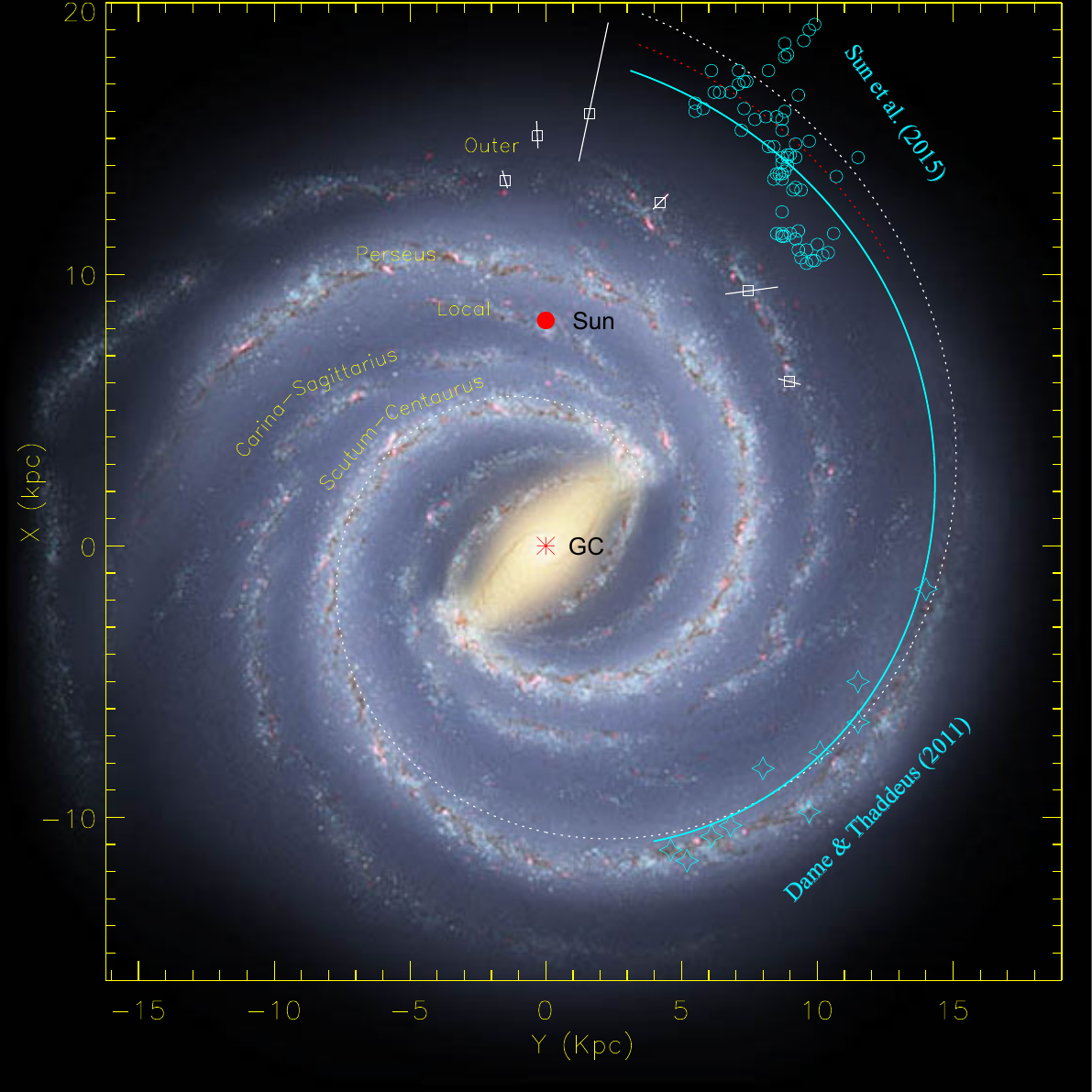}
\caption{
Locations of 72 selected molecular clouds in the far outer Galaxy (marked by cyan circles).
Locations of the high-mass star forming regions associated with the Outer Arm are denoted by white squares, with crossing white lines indicating distance uncertainties \citep{Reid2014}.
The white dashed line traces a log spiral with a mean pitch angle of $12^\circ$ that was fit to the Scutum–Centaurus arm. 
The red dashed line traces the log-periodic spiral that is fitting results to clouds ($120^\circ$\,<\,$l$\,<\,$150^\circ$) detected by \citet{Sun2015a}.
The cyan solid line traces the log-periodic spiral fitting results to clouds detected by \citet{Dame2011} and \citet{Sun2015a}.
Image adapted from \citet{Sun2015a}.}
\label{fig:target}
\end{figure}

We selected 72 Galactic edge clouds with masses on the order of $10^2$--$10^4$\,M$_{\odot}$ spanning the Galactic longitude range of $100^\circ$\,<\,$l$\,<\,$150^\circ$ from the Milky Way Imaging Scroll Painting (MWISP) project\footnote{\tiny \url{http://www.radioast.nsdc.cn/english/mwisp.php}} (see Fig.\,\ref{fig:target}; \citealt{Sun2015a}). 
This project is an unbiased northern Galactic plane CO survey conducted using the Delingha 13.7\,m telescope \citep{Su2019}. 
These Galactic edge clouds exhibit a wide range of Galactocentric radii, spanning approximately 14 to 22\,kpc, with a median equivalent radius of 17\,kpc (kinematic distance; \citealt{Sun2015a}). 
Typical radial velocity of these Galactic edge clouds is --102\,$\pm$\,10\,km\,s$^{-1}$.
Masses and sizes of these Galactic edge clouds are relatively small compared to the molecular clouds in the inner Galaxy \citep{Solomon1987,Heyer2009}. 
The H$_2$ column density of the dense clumps in these Galactic edge clouds ranges from 5$\times$10$^{20}$ to 3.6\,$\times$\,10$^{21}$\,cm$^{-2}$. 
While over ten clouds clearly show $^{13}$CO\,(1--0), a considerable number of other clouds were only marginally detected in this line, lacking a notable C$^{18}$O\,(1--0) feature \citep{Sun2015b}.
The structure of nine molecular clouds in these poorly studied subsolar metallicity regions and their relation with star formation 
have been investigated using the dense gas tracers HCN\,(1--0) and HCO$^+$\,(1--0) \citep{Braine2023}. Evidence of star formation within Galactic edge clouds is apparent \citep{Sun2015b}. 
Out of their 26 molecular clouds, a correlation with young stellar objects 
(YSOs) has been identified. Within this subset, 34 molecular cores are associated with YSOs. Furthermore, stellar clusters have been 
discovered in Digel Cloud 1 and 2  (e.g., \citealt{Yasui2006,Ruffle2007,Kobayashi2008,Izumi2014}), specifically in G131.016+1.524, 
G131.157+1.390, G137.759--0.983, and G137.775--1.067. These findings suggest widespread star formation even within the extreme outer Galaxy. 

\setcounter{figure}{1} 
\begin{figure*}[h]
\centering
\includegraphics[width=0.83\textwidth]{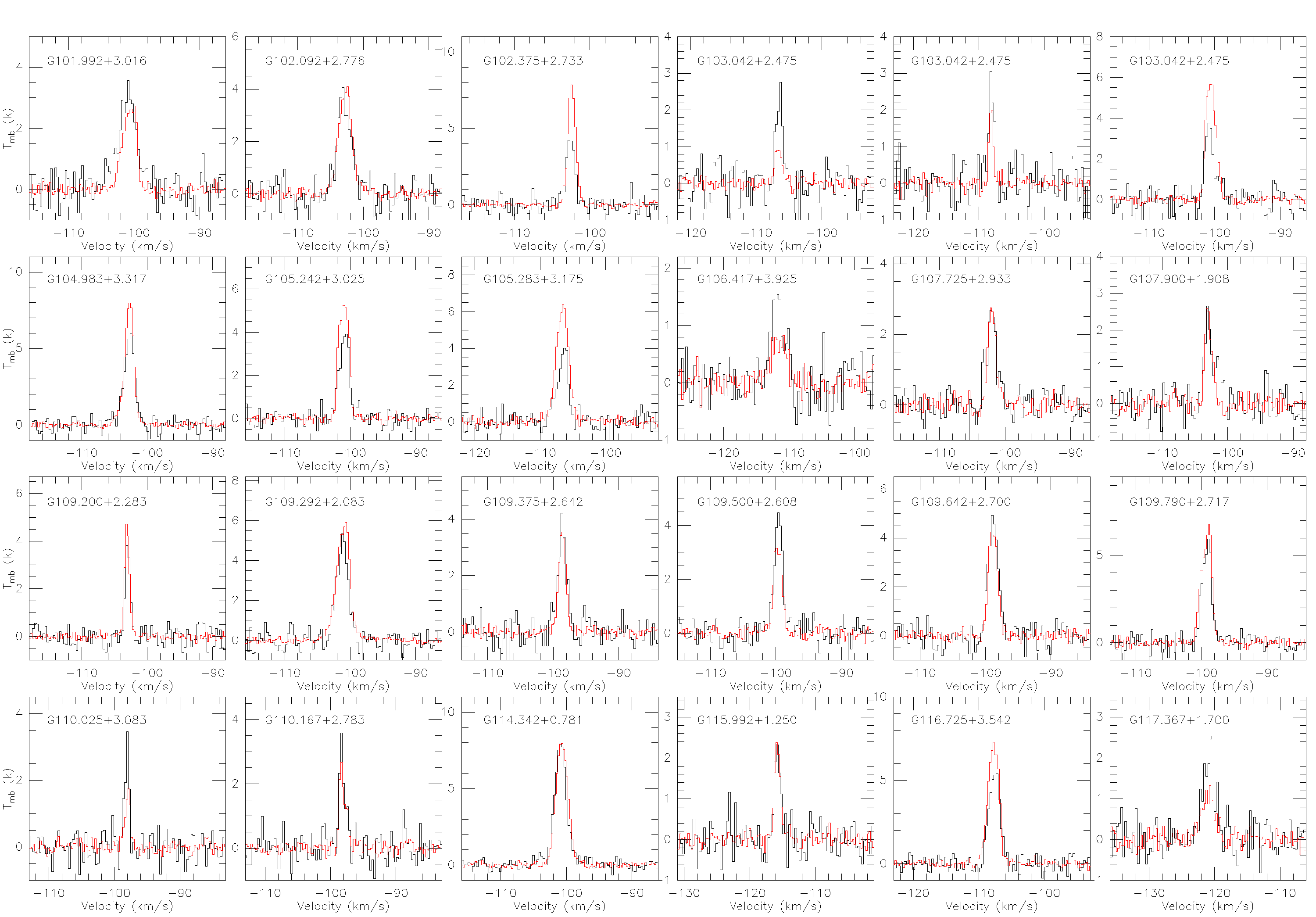}
\includegraphics[width=0.83\textwidth]{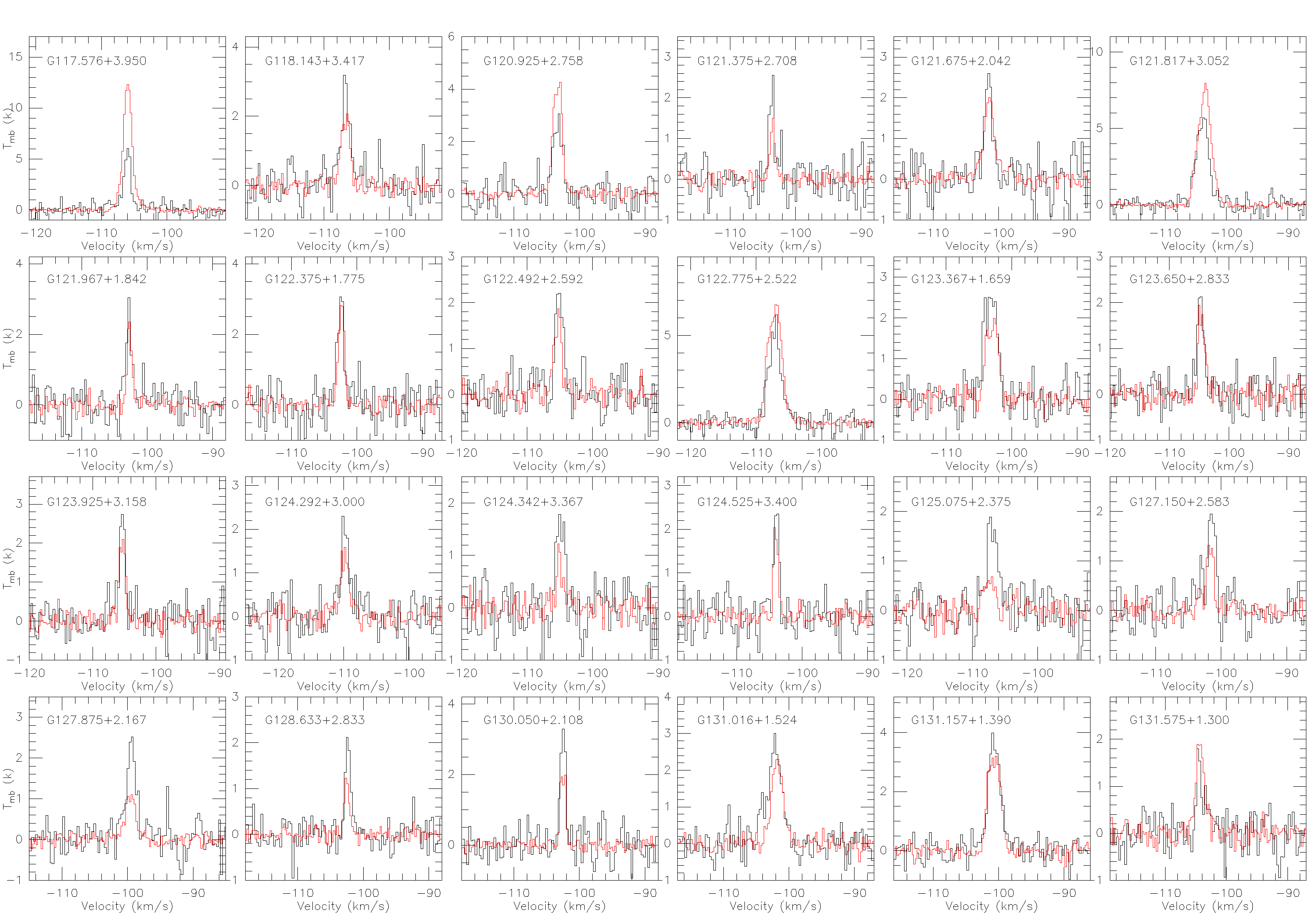}
\caption{The observed CO\,$J$\,=\,1--0 spectra (\emph{black line}) and 2--1 profiles (\emph{red line}) toward the Galactic edge clouds. 
The CO lines were extracted from the CO\,(1--0) emission peaks. 
The CO\,(2--1) line cubes have been smoothed to match that of the CO\,(1--0) line, 52$^{\prime\prime}$, for comparison.}
\label{fig:Spectra of CO(2--1)}
\end{figure*}

\setcounter{figure}{1}
\begin{figure*}[h]
\centering
\includegraphics[width=0.83\textwidth]{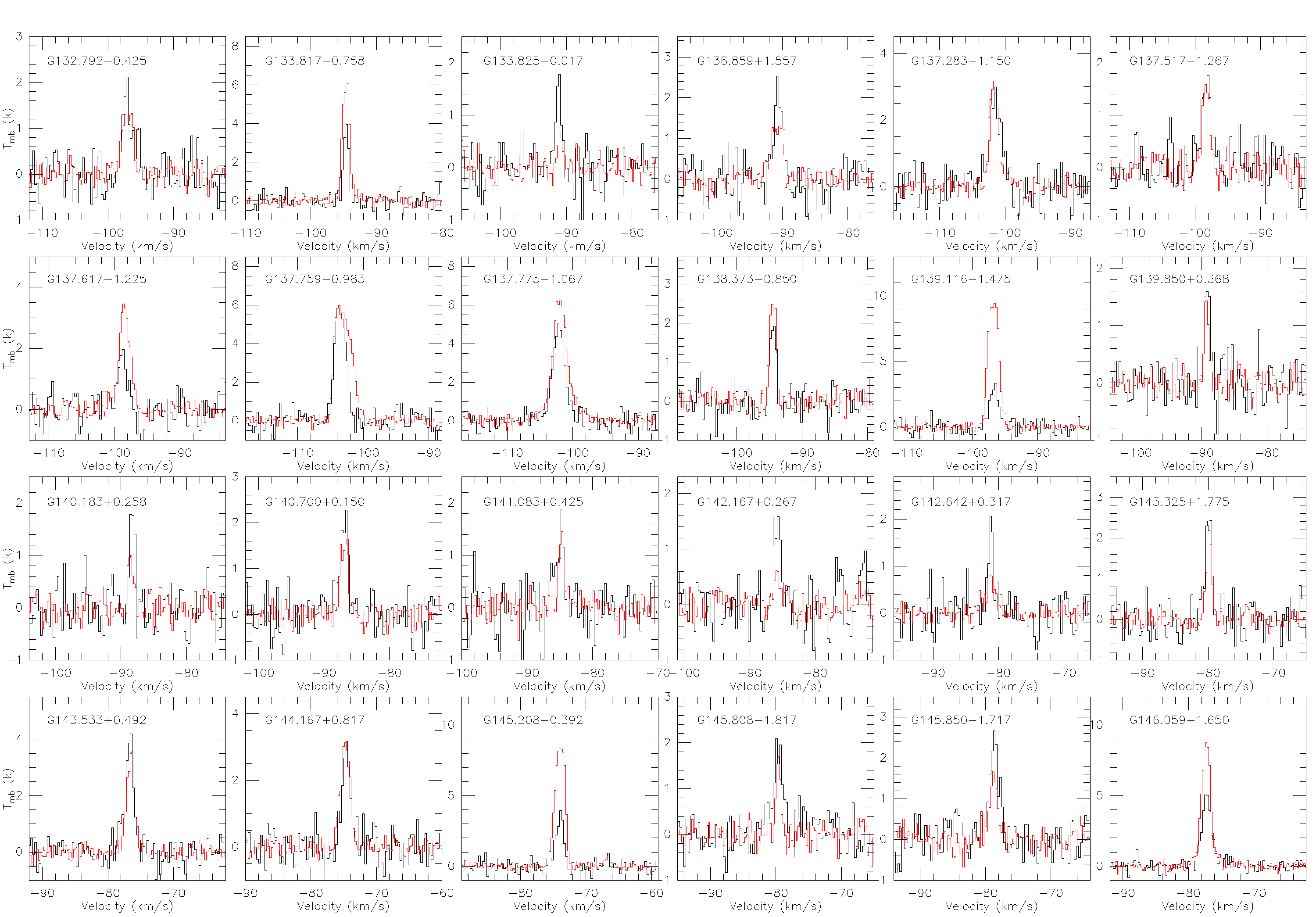}
\caption{continued.}
\end{figure*}

\subsection{Observations of CO\,(2--1)}
\label{sect:Observations}
The observations were conducted using the Institut de Radioastronomie Millimétrique 30\,m diameter (IRAM\,30\,m) telescope\footnote{\tiny Based on observations obtained with the IRAM\,30\,m telescope. IRAM is supported by INSU/CNRS (France), MPG (Germany), and IGN (Spain).} at Pico Veleta Observatory. 
Sky opacity at the telescope site varied from 0.2 to 0.4 at 2\,mm, with fluctuations observed over the course of the measurements\footnote{\tiny \url{https://publicwiki.iram.es/TelescopeSystemStatus}}. The pointing accuracy of the telescope was better than 3\,$^{\prime\prime}$. Pointing checks were conducted every 1 to 2 hours to ensure precision. 
Our CO\,(2--1) observations were carried out in December 2015, January 2016, and April 2016 utilizing the multibeam HEterodyne Receiver Array (HERA). 
HERA possesses the capability to efficiently observe nine positions on the sky simultaneously with dual-polarization. 
At a central frequency of $\sim$\,230\,GHz, the main parameters include a Half Power Beam Width (HPBW) of $\sim$\,11$^{\prime\prime}$, a main beam efficiency of $\sim$\,0.59, a forward efficieny of $\sim$\,0.92, and a system noise temperature of $\sim$\,500\,K ($T_{\rm A}^*$ scale). 
The backend spectrometer, with 5377 spectral channels and an aggregate bandwidth of $\sim$\,1\,GHz, provides a frequency resolution of 200\,kHz, corresponding to a channel width of $\sim$\,0.25\,km s$^{-1}$ at a frequency of $\sim$\,230\,GHz. 
We employed the On-The-Fly (OTF) observing mode to generate seventy-two maps ($\sim$\,100$^{\prime\prime}$\,$\times$\,100$^{\prime\prime}$) with steps of 3\hbox{$\,.\!\!^{\prime\prime}$}5 in both right ascension and declination. 
The central positions of the CO\,(2--1) observations in the 72 Galactic edge clouds are determined by identifying peaks in the CO\,(1--0) emission from the Delingha 13.7\,m telescope (see Fig.\,\ref{fig:Typical molecular structure}; \citealt{Sun2015a}).
The typical root-mean-square (RMS) noise levels for the CO\,(2--1) data are $\sim$\,0.3--0.5\,K ($T_{\rm mb}$ scale) at a channel width of $\sim$\,0.25\,km s$^{-1}$. 
The observed sources are listed in Table\,\ref{table:CO spectral parameters}.

\subsection{Archival data of CO\,(1--0)} 
CO\,(1--0) data were procured as part of the MWISP project \citep{Sun2015a}. The Delingha 13.7\,m telescope is equipped with a 3\,$\times$\,3 multibeam sideband-separation Superconducting Spectroscopic Array Receiver (SSAR), operating within a frequency range of 85--115\,GHz \citep{Shan2012}. 
The average zenith opacity is around 0.2 at 115\,GHz\footnote{\tiny \url{http://english.dlh.pmo.cas.cn/fs/}}. 
The typical system temperature for CO\,(1--0) is 250--300\,K ($T_{\rm A}^*$ scale). 
The pointing accuracy is better than 5$^{\prime\prime}$, while the tracking error is approximately 1$^{\prime\prime}$--3$^{\prime\prime}$. 
The MWISP observations encompass simultaneous measurements of $^{12}$CO, $^{13}$CO, and C$^{18}$O\,(1--0) lines, utilizing the OTF mode. 
The sampling interval is set at 15$^{\prime\prime}$, which is equivalent to the scan rate (50$^{\prime\prime}$\,s$^{-1}$) multiplied by the dump time (0.3\,s) in observations. 
The HPBW, the main beam efficiency, and the typical RMS noise level for CO\,(1--0) are approximately 52$^{\prime\prime}$, 0.46, and 0.5\,K, respectively. 
The backend instrumentation, utilizing a Fast Fourier Transform Spectrometer (FFTS), is configured with a 1\,GHz bandwidth and 16384 spectral channels, yielding a spectral resolution of 61\,kHz and a channel width of $\sim$\,0.16\,km\,s$^{-1}$ at 115\,GHz. 
The raw data were subsequently regridded into 30$^{\prime\prime}$\,$\times$\,30$^{\prime\prime}$ pixels, constituting a 30$^{\prime}$\,$\times$\,30$^{\prime}$ mosaic fits cube \citep{Su2019,Ma2021}. 

\subsection{Data reduction} 
\label{sect:data reduction} 
The spectral lines were processed utilizing the GILDAS\footnote{\tiny \url{http://www.iram.fr/IRAMFR/GILDAS}} package.
As a first step in this process, we removed low-quality CO\,(2--1) spectral lines and merged dual-polarization spectral lines to 
enhance the signal-to-noise ratio (S/N). The CO\,(1--0) spectral lines were smoothed over two consecutive channels to achieve a
velocity resolution of $\sim$\,0.32\,km\,s$^{-1}$, which closely aligns with the CO\,(2--1) channel width of $\sim$\,0.25\,km s$^{-1}$.
In order to match the spatial resolution of the CO\,(1--0) data, we smoothed the CO\,(2--1) data cube to a 52$^{\prime\prime}$
resolution and resampled the spectral lines at $\sim$\,30$^{\prime\prime}$ intervals. 
The parameters of the CO\,(2--1) spectra were only determined by aligning them with positions in the CO\,(1--0) peak emission and employing a Gaussian line profile for fitting, including the local standard of rest velocity ($V_{\rm LSR}$), main beam brightness peak temperature ($T_{\rm mb}$), 
full width at half maximum linewidth ($\Delta V$), and velocity-integrated intensity ($\int$$T_{\rm mb}$d$v$). These fitted results 
are listed in Table\,\ref{table:CO spectral parameters}. 
The parameters for the CO\,(1--0) spectra are detailed in Table\,1 of \citet{Sun2015a}.

Additionally, the heliocentric distance ($d$) and the Galactocentric distance ($R_{\rm g}$) were derived from the Galactic rotation curve 
as outlined by \citet{Reid2014}. The heliocentric distances of these Galactic edge clouds vary between approximately 9 and 15\,kpc, 
with a mean value of $\sim$\,11\,kpc \citep{Sun2015a}. Correspondingly, the spatial linear scales range from 2.3 to 3.7\,pc at a beam size 
of 52$''$, with an average of $\sim$\,2.8\,pc. In this work, the CO\,$J$=2--1/1--0 integrated intensity ratio 
($R_{21}$=$I_{\rm CO\,(2-1)}$/$I_{\rm CO\,(1-0)}$) was obtained for the Galactic edge clouds. 
Velocity ranges utilized for spectral integration are consistent across both CO transitions.
All CO\,(1--0) main beam brightness temperatures 
($T_{\rm mb}$) presented in this work have been calibrated for main beam efficiencies ($\eta_{\rm mb}$\,$\sim$\,0.46), being expressed as 
$T_{\rm mb}$\,=\,$T_{\rm A}^{*}$/$\eta_{\rm mb}$, where $T_{\rm A}^{*}$ represents the antenna temperature. 
CO\,(2--1) main beam brightness temperatures have been calibrated for main beam efficiencies ($B_{\rm eff}$\,$\sim$\,0.59) 
and the forward hemisphere efficiency ($F_{\rm eff}$\,$\sim$\,0.92), being expressed as 
$T_{\rm mb}$\,=\,($F_{\rm eff}/B_{\rm eff}$)$T_{\rm A}^{*}$. 

\setcounter{table}{0} 
\begin{table*}[htb]
\scriptsize
\caption{CO spectral parameters.}
\centering
\begin{tabular}
{ccccccccccccc}
\hline\hline
Sources & $l$ & $b$ & $d$  & $V_{\rm LSR}$ & $\Delta V$ & $T_{\rm mb}$ & $I_{\rm CO(2-1)}$ & $I_{\rm CO(1-0)}$ & $R_{21}$ &Structure &SF \\
& $^\circ$ & $^\circ$ & kpc & km\,s$^{-1}$ & km\,s$^{-1}$ & K & K\,km\,s$^{-1}$ & K\,km\,s$^{-1}$ & & & Activity \\
\hline
G101.992+3.016 &101.992 &3.016  &9.8  &-100.7\,$\pm$\,0.1 &2.4\,$\pm$\,0.1 &2.8 \,$\pm$\,0.3 &7.1 \,$\pm$\,0.2 &10.4\,$\pm$\,0.8 &0.68\,$\pm$\,0.06 &Moderate &With    \\
G102.092+2.776 &102.092 &2.776  &10.1 &-102.7\,$\pm$\,0.1 &2.4\,$\pm$\,0.1 &4.0 \,$\pm$\,0.2 &10.1\,$\pm$\,0.2 &8.9 \,$\pm$\,0.6 &1.13\,$\pm$\,0.08 &Compact  &With    \\
G102.375+2.733 &102.375 &2.733  &10.0 &-102.4\,$\pm$\,0.1 &1.2\,$\pm$\,0.1 &7.9 \,$\pm$\,0.1 &10.2\,$\pm$\,0.1 &6.2 \,$\pm$\,0.5 &1.65\,$\pm$\,0.14 &Compact  &With    \\
G103.042+2.475 &103.042 &2.475  &10.5 &-106.6\,$\pm$\,0.1 &1.3\,$\pm$\,0.2 &1.0 \,$\pm$\,0.1 &1.3 \,$\pm$\,0.1 &3.3 \,$\pm$\,0.4 &0.39\,$\pm$\,0.06 &Moderate &Without \\
G103.458+3.300 &103.458 &3.300  &10.7 &-108.1\,$\pm$\,0.1 &0.7\,$\pm$\,0.1 &2.2 \,$\pm$\,0.2 &1.6 \,$\pm$\,0.1 &2.8 \,$\pm$\,0.4 &0.57\,$\pm$\,0.09 &Moderate &Without \\
G103.729+2.867 &103.729 &2.867  &9.7  &-100.6\,$\pm$\,0.1 &1.8\,$\pm$\,0.1 &5.9 \,$\pm$\,0.4 &11.1\,$\pm$\,0.2 &6.1 \,$\pm$\,0.6 &1.82\,$\pm$\,0.19 &Compact  &With    \\
G104.983+3.317 &104.983 &3.317  &9.9  &-102.7\,$\pm$\,0.1 &1.6\,$\pm$\,0.1 &8.0 \,$\pm$\,0.4 &13.7\,$\pm$\,0.2 &11.4\,$\pm$\,0.5 &1.20\,$\pm$\,0.06 &Compact  &With    \\
G105.242+3.025 &105.242 &3.025  &9.6  &-101.0\,$\pm$\,0.1 &1.8\,$\pm$\,0.1 &5.7 \,$\pm$\,0.4 &10.7\,$\pm$\,0.2 &7.5 \,$\pm$\,0.4 &1.43\,$\pm$\,0.09 &Compact  &With    \\
G105.283+3.175 &105.283 &3.175  &10.4 &-106.6\,$\pm$\,0.1 &2.2\,$\pm$\,0.1 &6.3 \,$\pm$\,0.2 &14.6\,$\pm$\,0.2 &8.1 \,$\pm$\,0.6 &1.80\,$\pm$\,0.14 &Compact  &With    \\
G106.417+3.925 &106.417 &3.925  &11.1 &-111.6\,$\pm$\,0.2 &2.8\,$\pm$\,0.3 &0.7 \,$\pm$\,0.2 &2.2 \,$\pm$\,0.2 &3.2 \,$\pm$\,0.5 &0.69\,$\pm$\,0.13 &Diffuse  &Possible\\
G107.725+2.933 &107.725 &2.933  &9.7  &-102.0\,$\pm$\,0.1 &1.4\,$\pm$\,0.1 &2.8 \,$\pm$\,0.2 &4.2 \,$\pm$\,0.2 &4.8 \,$\pm$\,0.4 &0.88\,$\pm$\,0.09 &Moderate &Without \\
G107.900+1.908 &107.900 &1.908  &9.7  &-103.0\,$\pm$\,0.1 &1.2\,$\pm$\,0.1 &2.5 \,$\pm$\,0.3 &3.3 \,$\pm$\,0.2 &4.5 \,$\pm$\,0.5 &0.73\,$\pm$\,0.10 &Moderate &Without \\
G109.200+2.283 &109.200 &2.283  &9.8  &-103.1\,$\pm$\,0.1 &0.9\,$\pm$\,0.1 &4.7 \,$\pm$\,0.1 &4.4 \,$\pm$\,0.1 &3.4 \,$\pm$\,0.3 &1.29\,$\pm$\,0.12 &Compact  &Without \\
G109.292+2.083 &109.292 &2.083  &9.5  &-101.0\,$\pm$\,0.1 &2.3\,$\pm$\,0.1 &5.8 \,$\pm$\,0.4 &14.4\,$\pm$\,0.2 &11.5\,$\pm$\,0.5 &1.25\,$\pm$\,0.06 &Compact  &With    \\
G109.375+2.642 &109.375 &2.642  &9.2  &-98.6 \,$\pm$\,0.1 &1.3\,$\pm$\,0.1 &3.7 \,$\pm$\,0.1 &5.0 \,$\pm$\,0.2 &5.9 \,$\pm$\,0.5 &0.85\,$\pm$\,0.08 &Moderate &With    \\
G109.500+2.608 &109.500 &2.608  &9.3  &-99.7 \,$\pm$\,0.1 &1.5\,$\pm$\,0.1 &3.3 \,$\pm$\,0.2 &5.2 \,$\pm$\,0.2 &7.4 \,$\pm$\,0.4 &0.70\,$\pm$\,0.05 &Moderate &Possible\\
G109.642+2.700 &109.642 &2.700  &9.2  &-98.9 \,$\pm$\,0.1 &1.6\,$\pm$\,0.1 &4.5 \,$\pm$\,0.3 &7.7 \,$\pm$\,0.2 &9.1 \,$\pm$\,0.5 &0.85\,$\pm$\,0.06 &Moderate &Possible\\
G109.790+2.717 &109.790 &2.717  &9.2  &-99.2 \,$\pm$\,0.1 &1.9\,$\pm$\,0.1 &6.4 \,$\pm$\,0.7 &13.1\,$\pm$\,0.2 &10.3\,$\pm$\,0.5 &1.27\,$\pm$\,0.07 &Compact  &Possible\\
G110.025+3.083 &110.025 &3.083  &9.1  &-98.0 \,$\pm$\,0.1 &1.0\,$\pm$\,0.1 &1.7 \,$\pm$\,0.2 &1.7 \,$\pm$\,0.2 &2.9 \,$\pm$\,0.3 &0.59\,$\pm$\,0.10 &Moderate &Without \\
G110.167+2.783 &110.167 &2.783  &9.1  &-98.2 \,$\pm$\,0.1 &1.2\,$\pm$\,0.1 &2.3 \,$\pm$\,0.4 &2.9 \,$\pm$\,0.2 &3.3 \,$\pm$\,0.4 &0.88\,$\pm$\,0.13 &Moderate &Possible\\
G114.342+0.781 &114.342 &0.781  &9.5  &-100.7\,$\pm$\,0.1 &2.1\,$\pm$\,0.1 &8.1 \,$\pm$\,0.2 &18.6\,$\pm$\,0.2 &18.0\,$\pm$\,0.4 &1.03\,$\pm$\,0.03 &Compact  &With    \\
G115.992+1.250 &115.992 &1.250  &11.9 &-115.8\,$\pm$\,0.1 &1.0\,$\pm$\,0.1 &2.3 \,$\pm$\,0.1 &2.5 \,$\pm$\,0.2 &2.6 \,$\pm$\,0.4 &0.96\,$\pm$\,0.17 &Moderate &Without \\
G116.725+3.542 &116.725 &3.542  &10.5 &-107.8\,$\pm$\,0.1 &1.9\,$\pm$\,0.1 &7.4 \,$\pm$\,0.4 &15.2\,$\pm$\,0.2 &11.3\,$\pm$\,0.5 &1.35\,$\pm$\,0.07 &Compact  &With    \\
G117.367+1.700 &117.367 &1.700  &13.0 &-121.0\,$\pm$\,0.1 &2.4\,$\pm$\,0.2 &1.1 \,$\pm$\,0.2 &2.9 \,$\pm$\,0.2 &5.1 \,$\pm$\,0.5 &0.57\,$\pm$\,0.07 &Diffuse  &Possible\\
G117.576+3.950 &117.576 &3.950  &10.3 &-106.0\,$\pm$\,0.1 &1.6\,$\pm$\,0.1 &12.4\,$\pm$\,0.4 &20.8\,$\pm$\,0.2 &7.7 \,$\pm$\,0.6 &2.70\,$\pm$\,0.22 &Compact  &With    \\
G118.143+3.417 &118.143 &3.417  &10.4 &-106.7\,$\pm$\,0.1 &1.6\,$\pm$\,0.1 &2.0 \,$\pm$\,0.2 &3.4 \,$\pm$\,0.2 &4.9 \,$\pm$\,0.6 &0.69\,$\pm$\,0.10 &Moderate &Without \\
G120.925+2.758 &120.925 &2.758  &10.1 &-103.3\,$\pm$\,0.1 &1.6\,$\pm$\,0.1 &4.4 \,$\pm$\,0.3 &7.5 \,$\pm$\,0.2 &4.8 \,$\pm$\,0.5 &1.56\,$\pm$\,0.17 &Compact  &Without \\
G121.375+2.708 &121.375 &2.708  &10.2 &-103.5\,$\pm$\,0.1 &0.9\,$\pm$\,0.1 &1.4 \,$\pm$\,0.1 &1.3 \,$\pm$\,0.1 &2.4 \,$\pm$\,0.4 &0.54\,$\pm$\,0.10 &Moderate &Without \\
G121.675+2.042 &121.675 &2.042  &9.9  &-101.4\,$\pm$\,0.1 &2.0\,$\pm$\,0.2 &1.9 \,$\pm$\,0.2 &4.0 \,$\pm$\,0.2 &3.9 \,$\pm$\,0.5 &1.03\,$\pm$\,0.15 &Compact  &Without \\
G121.817+3.052 &121.817 &3.052  &10.3 &-103.6\,$\pm$\,0.1 &2.6\,$\pm$\,0.1 &7.5 \,$\pm$\,0.3 &20.3\,$\pm$\,0.2 &13.7\,$\pm$\,0.5 &1.48\,$\pm$\,0.06 &Compact  &With    \\
G121.967+1.842 &121.967 &1.842  &10.1 &-102.7\,$\pm$\,0.1 &0.8\,$\pm$\,0.1 &2.4 \,$\pm$\,0.1 &2.1 \,$\pm$\,0.2 &3.0 \,$\pm$\,0.5 &0.70\,$\pm$\,0.14 &Diffuse  &With    \\
G122.375+1.775 &122.375 &1.775  &10.1 &-102.3\,$\pm$\,0.1 &1.0\,$\pm$\,0.1 &3.0 \,$\pm$\,0.2 &3.3 \,$\pm$\,0.2 &3.9 \,$\pm$\,0.5 &0.85\,$\pm$\,0.12 &Diffuse  &Possible\\
G122.492+2.592 &122.492 &2.592  &10.6 &-105.3\,$\pm$\,0.1 &1.4\,$\pm$\,0.2 &1.7 \,$\pm$\,0.3 &2.5 \,$\pm$\,0.2 &3.5 \,$\pm$\,0.4 &0.71\,$\pm$\,0.10 &Diffuse  &Possible\\
G122.775+2.522 &122.775 &2.522  &11.0 &-107.2\,$\pm$\,0.1 &2.5\,$\pm$\,0.1 &6.7 \,$\pm$\,0.4 &17.9\,$\pm$\,0.2 &12.8\,$\pm$\,0.6 &1.40\,$\pm$\,0.07 &Compact  &Possible\\
G123.367+1.659 &123.367 &1.659  &10.4 &-103.0\,$\pm$\,0.1 &2.2\,$\pm$\,0.1 &2.0 \,$\pm$\,0.3 &4.5 \,$\pm$\,0.2 &6.5 \,$\pm$\,0.5 &0.69\,$\pm$\,0.07 &Moderate &With    \\
G123.650+2.833 &123.650 &2.833  &10.7 &-104.6\,$\pm$\,0.1 &1.3\,$\pm$\,0.1 &1.8 \,$\pm$\,0.3 &2.5 \,$\pm$\,0.2 &3.6 \,$\pm$\,0.4 &0.69\,$\pm$\,0.10 &Diffuse  &Possible\\
G123.925+3.158 &123.925 &3.158  &10.8 &-105.3\,$\pm$\,0.1 &1.2\,$\pm$\,0.1 &2.2 \,$\pm$\,0.3 &2.7 \,$\pm$\,0.2 &3.2 \,$\pm$\,0.4 &0.84\,$\pm$\,0.13 &Diffuse  &Possible\\
G124.292+3.000 &124.292 &3.000  &11.7 &-109.8\,$\pm$\,0.1 &1.6\,$\pm$\,0.2 &1.4 \,$\pm$\,0.3 &2.4 \,$\pm$\,0.2 &3.6 \,$\pm$\,0.5 &0.67\,$\pm$\,0.11 &Diffuse  &Without \\
G124.342+3.367 &124.342 &3.367  &10.8 &-105.1\,$\pm$\,0.1 &0.9\,$\pm$\,0.3 &1.1 \,$\pm$\,0.3 &1.1 \,$\pm$\,0.2 &2.8 \,$\pm$\,0.4 &0.39\,$\pm$\,0.10 &Diffuse  &Without \\
G124.525+3.400 &124.525 &3.400  &10.6 &-104.0\,$\pm$\,0.1 &0.8\,$\pm$\,0.1 &2.1 \,$\pm$\,0.3 &1.8 \,$\pm$\,0.1 &2.3 \,$\pm$\,0.3 &0.78\,$\pm$\,0.12 &Diffuse  &Without \\
G125.075+2.375 &125.075 &2.375  &11.3 &-107.4\,$\pm$\,0.2 &1.9\,$\pm$\,0.3 &0.6 \,$\pm$\,0.2 &1.3 \,$\pm$\,0.2 &3.9 \,$\pm$\,0.5 &0.33\,$\pm$\,0.07 &Diffuse  &Possible\\
G127.150+2.583 &127.150 &2.583  &10.6 &-101.7\,$\pm$\,0.1 &1.3\,$\pm$\,0.1 &1.4 \,$\pm$\,0.2 &1.8 \,$\pm$\,0.2 &3.1 \,$\pm$\,0.4 &0.58\,$\pm$\,0.10 &Diffuse  &Without \\
G127.875+2.167 &127.875 &2.167  &10.4 &-99.5 \,$\pm$\,0.1 &1.8\,$\pm$\,0.2 &1.1 \,$\pm$\,0.2 &2.1 \,$\pm$\,0.2 &4.9 \,$\pm$\,0.5 &0.43\,$\pm$\,0.06 &Moderate &With    \\
G128.633+2.833 &128.633 &2.833  &11.1 &-102.5\,$\pm$\,0.1 &0.9\,$\pm$\,0.1 &1.2 \,$\pm$\,0.1 &1.2 \,$\pm$\,0.1 &2.2 \,$\pm$\,0.3 &0.55\,$\pm$\,0.09 &Diffuse  &Without \\
G130.050+2.108 &130.050 &2.108  &11.4 &-102.5\,$\pm$\,0.1 &1.0\,$\pm$\,0.1 &2.2 \,$\pm$\,0.3 &2.2 \,$\pm$\,0.1 &3.4 \,$\pm$\,0.4 &0.65\,$\pm$\,0.09 &Diffuse  &With    \\
G131.016+1.524 &131.016 &1.524  &11.6 &-101.8\,$\pm$\,0.1 &2.0\,$\pm$\,0.1 &2.3 \,$\pm$\,0.2 &5.0 \,$\pm$\,0.2 &7.2 \,$\pm$\,0.5 &0.69\,$\pm$\,0.06 &Moderate &With    \\
G131.157+1.390 &131.157 &1.390  &11.3 &-100.7\,$\pm$\,0.1 &2.2\,$\pm$\,0.1 &3.3 \,$\pm$\,0.3 &7.8 \,$\pm$\,0.2 &8.6 \,$\pm$\,0.5 &0.91\,$\pm$\,0.06 &Moderate &With    \\
G131.575+1.300 &131.575 &1.300  &12.4 &-104.3\,$\pm$\,0.1 &1.6\,$\pm$\,0.2 &1.9 \,$\pm$\,0.2 &3.3 \,$\pm$\,0.2 &1.6 \,$\pm$\,0.4 &2.06\,$\pm$\,0.54 &Moderate &Without \\
G132.792-0.425 &132.792 &-0.425 &11.0 &-97.0 \,$\pm$\,0.1 &1.9\,$\pm$\,0.2 &1.4 \,$\pm$\,0.2 &2.9 \,$\pm$\,0.2 &3.9 \,$\pm$\,0.6 &0.74\,$\pm$\,0.13 &Moderate &Without \\
G133.817-0.758 &133.817 &-0.758 &10.7 &-94.6 \,$\pm$\,0.1 &1.3\,$\pm$\,0.1 &6.5 \,$\pm$\,0.4 &8.8 \,$\pm$\,0.2 &5.4 \,$\pm$\,0.3 &1.63\,$\pm$\,0.10 &Compact  &With    \\
G133.825-0.017 &133.825 &-0.017 &10.0 &-90.9 \,$\pm$\,0.1 &0.9\,$\pm$\,0.2 &0.7 \,$\pm$\,0.1 &0.6 \,$\pm$\,0.2 &1.8 \,$\pm$\,0.3 &0.33\,$\pm$\,0.13 &Diffuse  &Without \\
G136.859+1.557 &136.859 &1.557  &10.7 &-90.9 \,$\pm$\,0.1 &1.9\,$\pm$\,0.2 &1.4 \,$\pm$\,0.3 &2.8 \,$\pm$\,0.2 &3.5 \,$\pm$\,0.5 &0.80\,$\pm$\,0.13 &Moderate &Without \\
G137.283-1.150 &137.283 &-1.150 &14.0 &-101.6\,$\pm$\,0.1 &1.5\,$\pm$\,0.1 &3.2 \,$\pm$\,0.2 &5.0 \,$\pm$\,0.2 &6.9 \,$\pm$\,0.6 &0.72\,$\pm$\,0.07 &Moderate &Without \\
G137.517-1.267 &137.517 &-1.267 &13.1 &-98.3 \,$\pm$\,0.1 &1.3\,$\pm$\,0.2 &1.6 \,$\pm$\,0.2 &2.2 \,$\pm$\,0.2 &3.2 \,$\pm$\,0.5 &0.69\,$\pm$\,0.13 &Diffuse  &Without \\
G137.617-1.225 &137.617 &-1.225 &13.2 &-98.4 \,$\pm$\,0.1 &1.9\,$\pm$\,0.1 &3.4 \,$\pm$\,0.2 &6.9 \,$\pm$\,0.2 &5.6 \,$\pm$\,0.5 &1.23\,$\pm$\,0.12 &Compact  &With    \\
G137.759-0.983 &137.759 &-0.983 &14.7 &-103.1\,$\pm$\,0.1 &2.9\,$\pm$\,0.1 &6.0 \,$\pm$\,0.7 &18.7\,$\pm$\,0.2 &17.6\,$\pm$\,0.5 &1.06\,$\pm$\,0.04 &Compact  &With    \\
G137.775-1.067 &137.775 &-1.067 &14.4 &-102.0\,$\pm$\,0.1 &2.7\,$\pm$\,0.1 &6.2 \,$\pm$\,0.3 &17.5\,$\pm$\,0.3 &12.2\,$\pm$\,0.5 &1.43\,$\pm$\,0.07 &Compact  &With    \\
G138.373-0.850 &138.373 &-0.850 &12.3 &-94.4 \,$\pm$\,0.1 &1.1\,$\pm$\,0.1 &2.7 \,$\pm$\,0.3 &3.2 \,$\pm$\,0.2 &3.2 \,$\pm$\,0.3 &1.00\,$\pm$\,0.12 &Moderate &Without \\
G139.116-1.475 &139.116 &-1.475 &13.4 &-96.8 \,$\pm$\,0.1 &1.9\,$\pm$\,0.1 &10.2\,$\pm$\,0.6 &20.4\,$\pm$\,0.2 &6.7 \,$\pm$\,0.6 &3.04\,$\pm$\,0.28 &Compact  &Without \\
G139.850+0.368 &139.850 &0.368  &11.4 &-89.3 \,$\pm$\,0.1 &0.7\,$\pm$\,0.1 &1.6 \,$\pm$\,0.1 &1.2 \,$\pm$\,0.2 &2.9 \,$\pm$\,0.3 &0.41\,$\pm$\,0.09 &Diffuse  &Without \\
G140.183+0.258 &140.183 &0.258  &11.4 &-88.5 \,$\pm$\,0.1 &0.9\,$\pm$\,0.2 &1.1 \,$\pm$\,0.2 &1.0 \,$\pm$\,0.2 &1.8 \,$\pm$\,0.3 &0.56\,$\pm$\,0.15 &Diffuse  &Without \\
G140.700+0.150 &140.700 &0.150  &11.2 &-86.9 \,$\pm$\,0.1 &1.4\,$\pm$\,0.2 &1.6 \,$\pm$\,0.3 &2.5 \,$\pm$\,0.2 &2.5 \,$\pm$\,0.5 &1.00\,$\pm$\,0.22 &Diffuse  &Without \\
G141.083+0.425 &141.083 &0.425  &10.8 &-84.7 \,$\pm$\,0.1 &0.9\,$\pm$\,0.2 &1.4 \,$\pm$\,0.2 &1.3 \,$\pm$\,0.2 &2.6 \,$\pm$\,0.5 &0.50\,$\pm$\,0.13 &Diffuse  &Without \\
G142.167+0.267 &142.167 &0.267  &11.6 &-85.8 \,$\pm$\,0.2 &1.2\,$\pm$\,0.6 &0.6 \,$\pm$\,0.1 &0.7 \,$\pm$\,0.3 &2.3 \,$\pm$\,0.4 &0.30\,$\pm$\,0.15 &Diffuse  &Without \\
G142.642+0.317 &142.642 &0.317  &10.5 &-81.3 \,$\pm$\,0.1 &1.7\,$\pm$\,0.3 &0.8 \,$\pm$\,0.2 &1.4 \,$\pm$\,0.2 &2.0 \,$\pm$\,0.3 &0.70\,$\pm$\,0.15 &Diffuse  &Possible\\
G143.325+1.775 &143.325 &1.775  &10.4 &-79.9 \,$\pm$\,0.1 &1.0\,$\pm$\,0.1 &2.4 \,$\pm$\,0.2 &2.5 \,$\pm$\,0.2 &3.1 \,$\pm$\,0.4 &0.81\,$\pm$\,0.13 &Diffuse  &With    \\
\hline\,
\end{tabular}
\label{table:CO spectral parameters}
\tablefoot{Column 1: source name. Columns 2 and 3: Galactic coordinates. Column 4: heliocentric distance taken from \citet{Sun2015a}. Columns 5–8: results of Gaussian fits to the CO\,$J$=2--1 spectra. Column 9: integrated intensities of the  CO\,$J$=1--0 spectra taken from \citet{Sun2015a}. Column 10: the CO\,$J$=2--1/1--0 integrated intensity ratio. 
Column 11: classification for the degree of compactness of molecular clouds. Classification by eye, according to velocity-integrated intensity maps of molecular clouds (see Fig.\,\ref{fig:Typical molecular structure}). 
Column 12: Galactic edge clouds with or without star-forming (SF) activity, which are determined by {\it WISE}-certified young stars according to \citet{Sun2015b}.}\\
\end{table*}

\setcounter{table}{0} 
\begin{table*}[htb]
\scriptsize
\caption{continued.}
\centering
\begin{tabular}
{ccccccccccccc}
\hline\hline
Sources & $l$ & $b$ & $d$  & $V_{\rm LSR}$ & $\Delta V$ & $T_{\rm mb}$ & $I_{\rm CO(2-1)}$ & $I_{\rm CO(1-0)}$ & $R_{21}$ &Structure &SF \\
& $^\circ$ & $^\circ$ & kpc & km\,s$^{-1}$ & km\,s$^{-1}$ & K & K\,km\,s$^{-1}$ & K\,km\,s$^{-1}$ & & & Activity \\
\hline
G143.533+0.492 &143.533 &0.492  &9.7  &-76.5 \,$\pm$\,0.1 &1.4\,$\pm$\,0.1 &3.5 \,$\pm$\,0.3 &5.2 \,$\pm$\,0.2 &6.5 \,$\pm$\,0.5 &0.80\,$\pm$\,0.07 &Compact  &With    \\
G144.167+0.817 &144.167 &0.817  &9.4  &-74.7 \,$\pm$\,0.1 &1.7\,$\pm$\,0.1 &3.1 \,$\pm$\,0.3 &5.7 \,$\pm$\,0.2 &4.6 \,$\pm$\,0.5 &1.24\,$\pm$\,0.15 &Compact  &Without \\
G145.208-0.392 &145.208 &-0.392 &9.7  &-73.9 \,$\pm$\,0.1 &1.7\,$\pm$\,0.1 &9.2 \,$\pm$\,0.7 &16.9\,$\pm$\,0.2 &8.8 \,$\pm$\,0.5 &1.92\,$\pm$\,0.12 &Compact  &Without \\
G145.808-1.817 &145.808 &-1.817 &11.6 &-79.6 \,$\pm$\,0.1 &0.8\,$\pm$\,0.1 &1.8 \,$\pm$\,0.1 &1.5 \,$\pm$\,0.2 &3.6 \,$\pm$\,0.5 &0.42\,$\pm$\,0.09 &Diffuse  &Without \\
G145.850-1.717 &145.850 &-1.717 &11.3 &-78.7 \,$\pm$\,0.1 &1.4\,$\pm$\,0.2 &1.5 \,$\pm$\,0.2 &2.4 \,$\pm$\,0.3 &4.4 \,$\pm$\,0.5 &0.55\,$\pm$\,0.10 &Diffuse  &Without \\
G146.059-1.650 &146.059 &-1.650 &11.0 &-77.3 \,$\pm$\,0.1 &1.6\,$\pm$\,0.1 &8.9 \,$\pm$\,0.4 &15.3\,$\pm$\,0.2 &8.7 \,$\pm$\,0.4 &1.76\,$\pm$\,0.09 &Compact  &With    \\
\hline\,
\end{tabular}
\end{table*}

\begin{figure*}[t]
\centering
\includegraphics[width=1.0\textwidth]{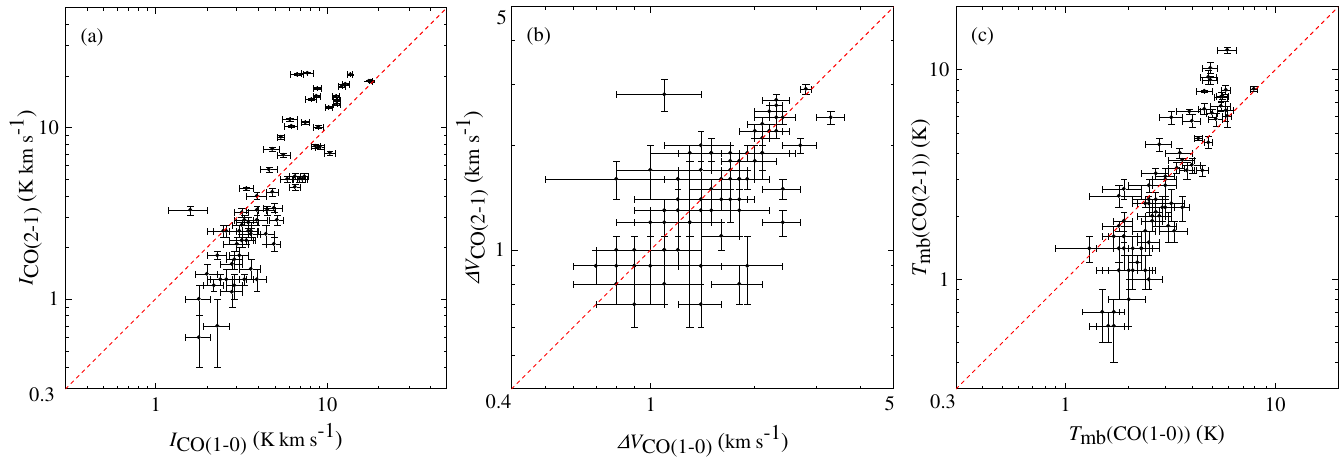}
\caption{Correlations of integrated intensities (a), linewidths (b), and peak temperatures (c) of CO\,(1--0) and (2--1) lines. The red dashed lines indicate Y\,=\,X.}
\label{fig:I-dV-Tmb}
\end{figure*}

\section{Results}
\label{sect:Results}
\subsection{Overview}
\label{sect:Overview}
The observed CO\,(1--0) and (2--1) spectra are depicted in Fig.\,\ref{fig:Spectra of CO(2--1)}. The CO\,(2--1) lines are detected 
in all sources. The integrated intensity distributions of CO\,(2--1) for the 72 Galactic edge clouds will be detailed in Luo et al. (in prep.). 
Among these observed molecular clouds, 25 present compact structures\footnote{\tiny The forthcoming work by Luo et al. (in prep) will provide a detailed account of the identification of "compact" CO clumps extracted using {\tt quickclump} \citep{Sidorin2017}. The angular sizes of these CO clumps range from approximately 16$^{\prime\prime}$ to 58$^{\prime\prime}$, with a mean size of $\sim$27$^{\prime\prime}$. In linear terms, these clumps span from 0.6 to 3.9\,pc, with an average size of $\sim$1.3\,pc.}, 25 display diffuse structures, and 22 manifest intermediate morphologies between the two based on identification by eye (see Table\,\ref{table:CO spectral parameters}). 
Typical examples of these three structures are shown in Fig.\,\ref{fig:Typical molecular structure}.
The fitted line parameters, integrated intensity, linewidth, and peak temperature of CO\,(1--0) and (2--1) are depicted in
Fig.\,\ref{fig:I-dV-Tmb}. Good correlations have been observed between these parameters. The centroid velocities for both
lines are similar (see Table\,\ref{table:CO spectral parameters} and Table\,1 in \citealt{Sun2015a}). The linewidths of both CO\,(1–0) and (2–1) are notably narrow,
exhibiting a similar range. For the CO\,(1–0) transition, the linewidth ranges from 0.7 to 3.3\,km\,s$^{-1}$, with an average of 1.6\,$\pm$\,0.3\,km\,s$^{-1}$ (errors given here and elsewhere are standard deviations of the mean).
Similarly, in the CO\,(2--1) transition, the linewidth ranges from 0.7 to 2.9\,km\,s$^{-1}$, with an average of 1.6\,$\pm$\,0.2\,km\,s$^{-1}$.
The agreement observed in both CO\,(1--0) and (2--1) lines indicates that they possibly trace similar components of molecular gas. 
In these Galactic edge clouds, $R_{21}$ therefore appears to be a meaningful quantity to be studied.

\subsection{$R_{21}$ line ratios}
\label{sect:Line Ratios of CO}
As illustrated in Fig.\,\ref{fig:Spectra of CO(2--1)} and Table\,\ref{table:CO spectral parameters},
approximately 40\% of the CO\,(1--0) emissions in these Galactic edge clouds are weaker than CO\,(2--1). 
Specifically, the observed range of the $R_{21}$ ratio within the Galactic edge clouds ranges from 0.30 to 3.04, with a mean of 0.97\,$\pm$\,0.12 and a median of 0.81\,$\pm$\,0.10 (see Table\,\ref{table:CO spectral parameters}). 
Our observations show significant variations in the $R_{21}$ ratios by a factor of around five, reaching a full order of magnitude when comparing the lowest with the highest $R_{21}$ value.

\begin{figure}[t]
\centering
\includegraphics[width=0.5\textwidth]{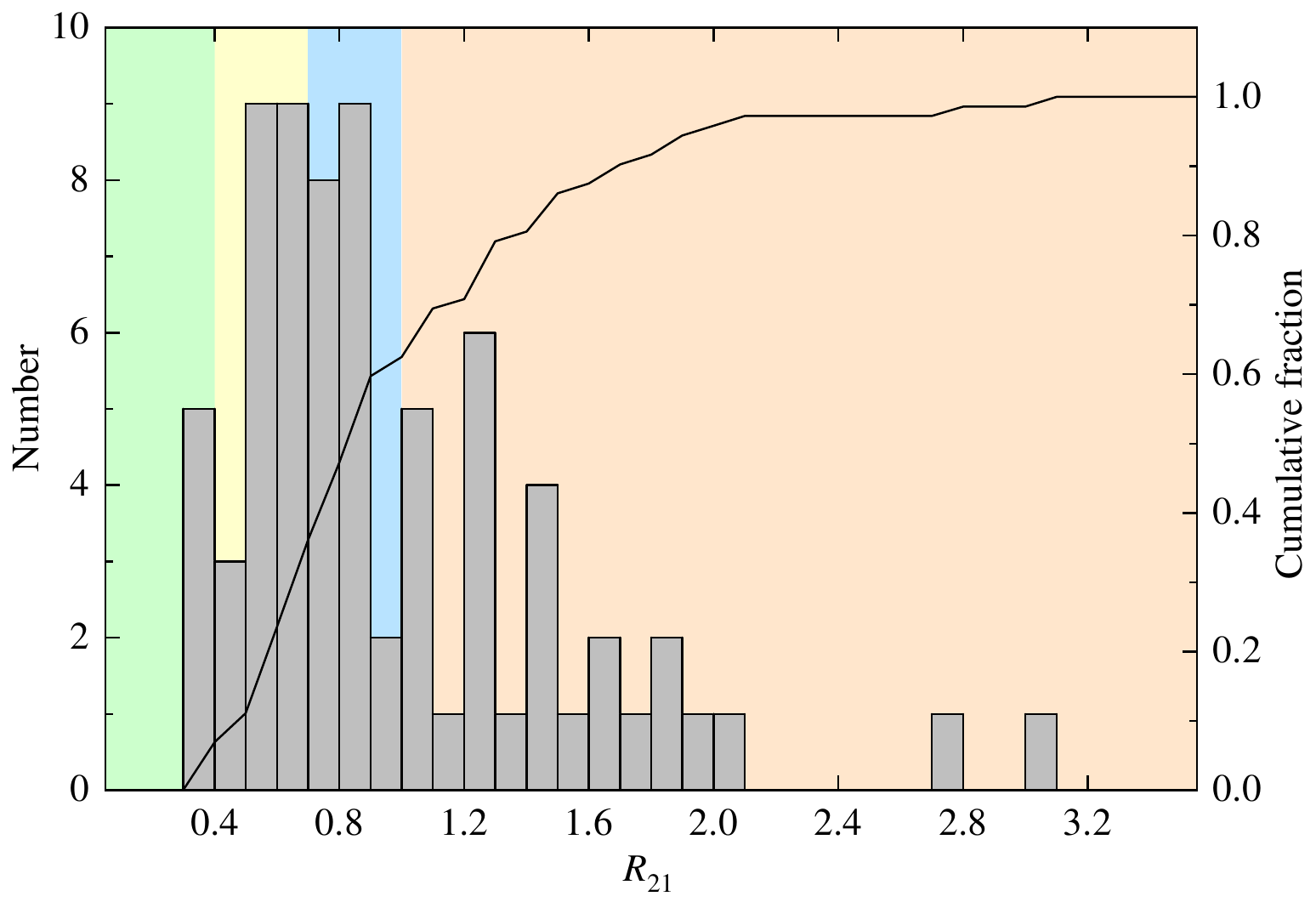}
\caption{Histogram of the $R_{21}$ ratios.
The solid line shows the cumulative distribution function.
The background colors green, yellow, cyan, and orange are used to denote the VLRG, LRG, HRG, and VHRG categories, respectively (see Sect.\,\ref{sect:CO-Line-Ratio}).}
\label{fig:histgram}
\end{figure}

The statistical histogram and cumulative distribution of the $R_{21}$ ratio are depicted in Fig.\,\ref{fig:histgram}. 
The majority of $R_{21}$ values fall within the range of 0.5--0.9 and there is a systematic prevalence of occurrences for $R_{21}$\,>\,0.7 compared to $R_{21}$\,<\,0.7. 
Additionally, there exist three sources where $R_{21}$ exceeds 2.0, 
which are G117.576+3.950, G131.575+1.300, and G139.116--1.475. Such elevated $R_{21}$ ratios stem from the 
considerably higher peak temperature of CO\,(2--1) relative to CO\,(1--0), with the exception of G131.575+1.300, 
where the linewidth is the primary contributor (see Fig.\,\ref{fig:Spectra of CO(2--1)} and discussion in 
Sect.\,\ref{Sect:Compact Clouds}). 
Counts of $R_{21}$ values categorized into four distinct groups (see Sect.\,\ref{sect:CO-Line-Ratio}) are presented in
Table\,\ref{table:counts}. The quantities of the VLRG, LRG, HRG, and VHRG in the 72 Galactic edge clouds are 5, 21, 19, 
and 27, respectively. Percentages of VLRG, LRG, HRG, and VHRG are 6.9$\%$, 29.2$\%$, 26.4$\%$, and 37.5$\%$, respectively, 
indicating an increasing trend. These results suggest that the proportion of high $R_{21}$ ratios ($\ge$0.7) is significant 
in molecular clouds at the edge of the Galaxy. Specifically, the mean $R_{21}$ values are as follows: 0.35\,$\pm$\,0.11 
with a corresponding median of 0.33\,$\pm$\,0.10, 0.59\,$\pm$\,0.10 with a median of 0.58\,$\pm$\,0.10, 0.80\,$\pm$\,0.11 
with a median of 0.80\,$\pm$\,0.12, and 1.51\,$\pm$\,0.14 with a median of 1.40\,$\pm$\,0.12 for VLRG, LRG, HRG, and VHRG, 
respectively (see Table\,\ref{table:counts}). For each of these four $R_{21}$ categories of molecular gas, both mean and median ratios 
are approximately equivalent. 

\begin{table}[t]
\scriptsize
\caption{Counts of four $R_{21}$ categories.}
\centering
\begin{tabular}
{cccccc}
\hline\hline
Categories& Range& Mean& Median& Counts& Percentage \\
\hline
VLRG &$R_{21} < 0.4$         &0.35\,$\pm$\,0.11 &0.33\,$\pm$\,0.10 &5  &6.9$\%$ \\
LRG  &$0.4 \le R_{21} < 0.7$ &0.59\,$\pm$\,0.10 &0.58\,$\pm$\,0.10 &21 &29.2$\%$\\
HRG  &$0.7 \le R_{21} < 1.0$ &0.80\,$\pm$\,0.11 &0.80\,$\pm$\,0.12 &19 &26.4$\%$\\
VHRG &$R_{21} \ge 1.0$       &1.51\,$\pm$\,0.14 &1.40\,$\pm$\,0.12 &27 &37.5$\%$\\
\hline
\end{tabular}
\label{table:counts}
\end{table}

\begin{figure*}[t]
\centering
\includegraphics[width=1.0\textwidth]{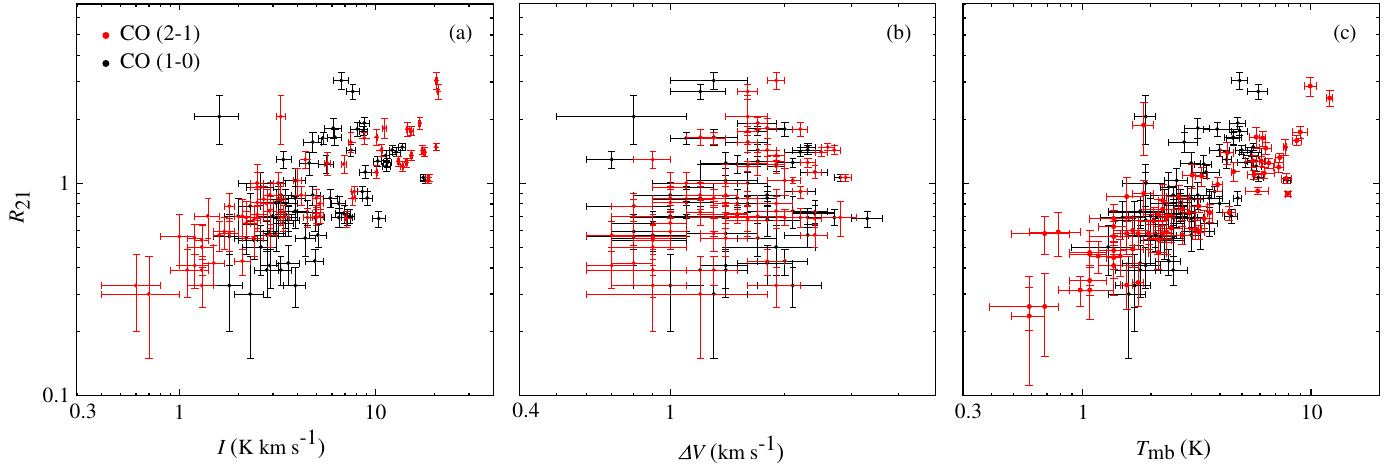}
\caption{Correlations between $R_{21}$ and integrated intensities (a), linewidths (b), and peak temperatures (c) of CO lines. Addressed are all our observed 72 Galactic edge clouds. The black and red points denote CO\,(1--0) and (2--1), respectively.} 
\label{fig:I-R} 
\end{figure*}

\subsection{Correlations of $R_{21}$ with CO line parameters} 
\label{Sec:Relation}
The correlations between the $R_{21}$ ratios and CO line parameters are illustrated in Fig.\,\ref{fig:I-R}. Positive correlations
are observed between the $R_{21}$ ratios and both the CO integrated intensities and peak temperatures. In general, the Galactic edge clouds exhibit higher $R_{21}$ in regions with higher CO intensities. 
There is a weak correlation between the $R_{21}$ ratios and the linewidths of the CO lines. 
Furthermore, the linewidth ratio is close to unity (see Sect.\,\ref{sect:Overview}), indicative of uniform linewidths in both CO\,(1--0) and CO\,(2--1). 
Consequently, the CO integrated intensity ratio aligns with the peak temperature ratio, as depicted in Fig.\,\ref{fig:R-R}.

\section{Discussion}
\label{sect:discussion}

\subsection{Variation of $R_{21}$ ratios in the Milky Way}
\label{Sec:Variation}
The variation of the $R_{21}$ ratio with Galactocentric distance across the entire Galactic disk is depicted in Fig.\,\ref{fig:Rg-R21}. 
Within the inner Galaxy, the $R_{21}$ ratio appears to decrease as a function of Galactocentric radius. Notably, the $R_{21}$ ratios decrease from 0.9--1.0 in the central region to $\sim$0.7 within the molecular ring of the Milky Way \citep{Chiar1994,Oka1998,Sawada2001}. 
The $R_{21}$ ratio varies from 0.75\,$\pm$\,0.06 at $R_{\rm g}$\,=\,4\,kpc to 0.6\,$\pm$\,0.1 in the solar neighborhood (e.g., \citealt{Handa1993,Chiar1994,Handa1999,Sakamoto1995,Sakamoto1997,Hasegawa1997a,Usuda1999}). 
However, within a range of approximately 8--14\,kpc, the $R_{21}$ ratio displays significant variability and does not adhere to this trend. 
In the inner Galaxy, a mixture of compact components containing high ratio gas and diffuse gas components enveloping these compact structures, characterized by lower ratios, is observed. In contrast, only compact components emitting CO are detected in the outer Galaxy. 
These results suggest that the relevant $R_{21}$ ratios in the outer Galaxy may be an indication of dense molecular gas poised for significant star formation \citep{Usuda1999}. 
These measurements of $R_{21}$ are consistent with those obtained from the Galactic edge clouds, suggesting that the physical conditions within these molecular clouds may be analogous in both the outer Galaxy and the Galactic edge clouds. 

\begin{figure}[t]
\centering
\includegraphics[width=0.48\textwidth]{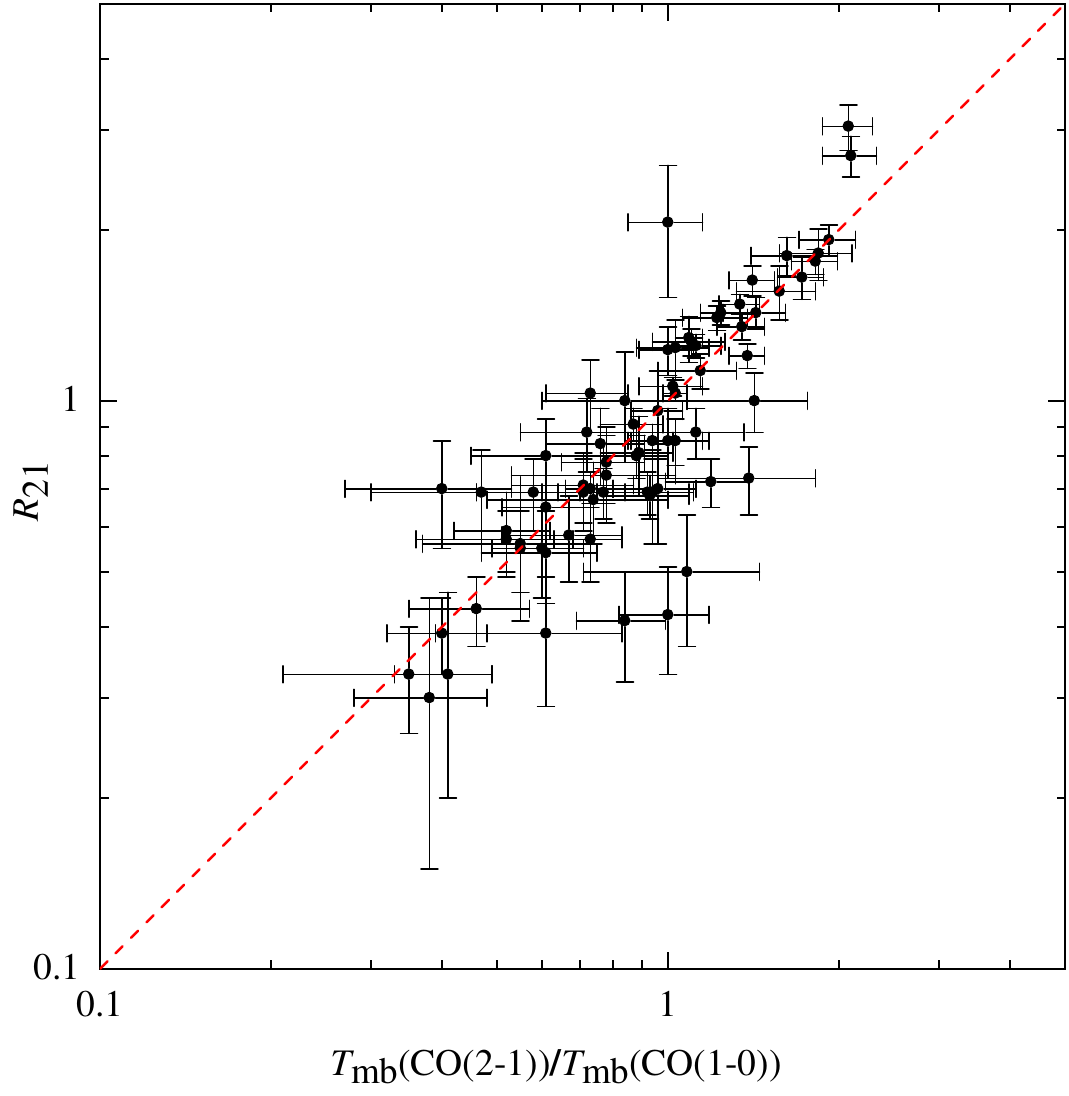}
\caption{Correlation between $R_{21}$ ratios and CO brightness temperature ratios. The red dashed line indicates Y\,=\,X.}
\label{fig:R-R} 
\end{figure}

\subsection{$R_{21}$ in external galaxies}
\subsubsection{Magellanic Clouds}
\label{sect:Magellanic Clouds}
The metallicity of the Galactic edge clouds exhibits similarities to metal-poor galaxies such as the LMC and SMC \citep{Yasui2006, Yasui2008}. 
The $R_{21}$ ratios in the Galactic edge clouds (see Sect.\,\ref{sect:Line Ratios of CO}) show a broader distribution than those in the LMC 
(0.5-->1.3; \citealt{Heikkila1999,Bolatto2000,Sorai2001}), likely due to a larger number of observed samples and higher linear resolution. 
The average $R_{21}$ value of 0.92\,$\pm$\,0.05 in the LMC \citep{Sorai2001} is similar to the mean $R_{21}$ in the Galactic edge clouds. 
The $R_{21}$ ratio within 30 Doradus cloud complex is 0.95\,$\pm$\,0.06, consistent with our average results in the Galactic edge clouds but higher than those observed in the outer regions of the LMC. 
The maximum $R_{21}$ value of 3.04 in the Galactic edge
clouds is consistent with an extended envelope with $R_{21}$\,$>$\,3 in the N159/N160 complex of the LMC \citep{Bolatto2000}. 
Subsequently, the presence of the VHRG with $R_{21}$ values exceeding 2.0 has been discovered within the N83/N84 molecular cloud complex of the SMC. 
The typical value of $R_{21}$ is around 1.0 in the molecular clouds of the SMC \citep{Heikkila1999,Bolatto2003}, which is also consistent with our mean $R_{21}$ observed in the Galactic edge clouds. 
The similar mean values of $R_{21}$ observed in the Galactic edge clouds, the LMC, 
and the SMC indicate that the physical conditions within these molecular clouds may be comparable.

In the LMC, a radial gradient in the $R_{21}$ ratio is observed at distances from the center out to $\sim$\,4\,kpc, 
with the ratio measuring 0.94\,$\pm$\,0.34 in the inner region ($\lesssim$\,2\,kpc from the kinematic center) and 0.69\,$\pm$\,0.27 in the outer region ($\gtrsim$\,2\,kpc from the center, excluding the 30\,Doradus complex) \citep{Sorai2001}. This indicates that the inner clouds are both warmer ($T_{\rm kin}$\,>\,20--40\,K) and/or denser ($n_{\rm H_2}$>$10^3$\,cm$^{-3}$) than the outer clouds of the LMC. 
The radial variation observed in the inner Galaxy exhibits a similar trend to that observed in the LMC, which may indicate a shared mechanism that dictates the average physical conditions of molecular gas on large scales within disk galaxies, despite their morphological differences \citep{Sorai2001}. 
However, the mean $R_{21}$ ratio in the Galactic edge clouds is higher compared to the outer region of the LMC, indicating potential differences in the physical conditions of molecular clouds between the edge of our Galaxy and the outer region of the LMC. 

\subsubsection{Other nearby galaxies}
\label{sect:Nearby Galaxy}
\begin{figure}[t]
\centering
\includegraphics[width=0.49\textwidth]{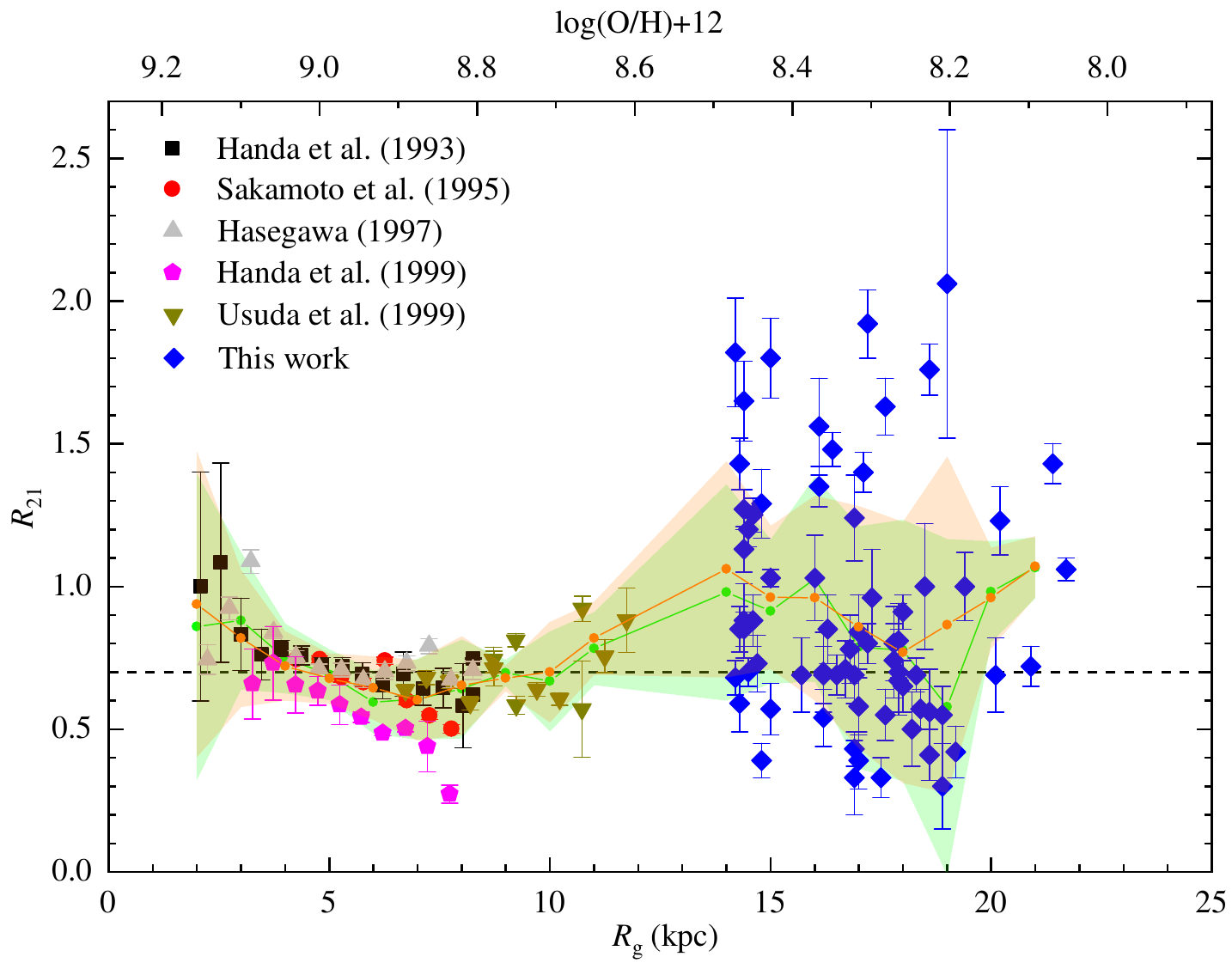}
\caption{Variation of the $R_{21}$ ratios with Galactocentric distance $R_{\rm g}$. 
The orange values refer to 1\,kpc bins with orange straight lines connecting them (unweighted mean value within a bin), and dispersion indicating the propagated uncertainty. 
The green values, weighted by the square of the uncertainties, correspond to 1\,kpc bins. 
Values of $R_{21}$\,>\,2.2 are not shown here and values of $R_{\rm g}$\,<\,14\,kpc are obtained from previous measurements \citep{Handa1993,Handa1999,Sakamoto1995,Hasegawa1997a,Usuda1999}. The black horizontal dotted line denotes a value of $R_{21}$\,=\,0.7. 
} 
\label{fig:Rg-R21}
\end{figure}

The investigation of the $R_{21}$ ratio has been conducted on a significant number of extragalactic sources located beyond the Magellanic Clouds. 
Analogous to the Milky Way and Magellanic Clouds, the $R_{21}$ ratio exhibits a variation ranging from approximately 0.3 to 2.6, as established through CO multi-line surveys of external galaxies 
(e.g., \citealt{Braine1992,Leroy2009,Papadopoulos2012,Leroy2013,Leroy2022,Israel2020,denBrok2021,Yajima2021,Keenan2024}). 
Similar to the situation in the Milky Way, the Galactic $R_{21}$ ratio displays a slight trend of an initial radial decline followed by a higher dispersion in the nearby barred spiral galaxy NGC\,2903 \citep{denBrok2021}. 
In NGC\,2903, high star formation efficiency concentrates in central zones and \ion{H}{II} regions distributed along its bar \citep{Alonso-Herrero2001, Popping2010}. 
Notably, the central regions of many of these galaxies display consistently elevated $R_{21}$ ratios (e.g., \citealt{Braine1992,Leroy2009,Leroy2013,Israel2020,denBrok2021,Yajima2021}). 
The $R_{21}$ dispersion increases with distance from the center, both for our Galaxy and nearby galaxies (see also Fig.\,\ref{fig:Rg-R21} and Fig.\,3 in \citealt{denBrok2021}). 

\begin{table*}[t]
\caption{The typical linewidth and brightness temperature of CO lines in different $R_{21}$ classifications}
\centering
\begin{tabular}
{lccccccccccc}
\hline\hline
& \multicolumn{5}{c}{CO\,(1--0)}& & \multicolumn{5}{c}{CO\,(2--1)}\\
\cline{2-6} \cline{8-12}
Categories& \multicolumn{2}{c}{$\Delta V$ (km\,s$^{-1}$)}& & \multicolumn{2}{c}{$T_{\rm mb}$ (K)}& & \multicolumn{2}{c}{$\Delta V$ (km\,s$^{-1}$)}& 
& \multicolumn{2}{c}{$T_{\rm mb}$ (K)}\\
\cline{2-3} \cline{5-6} \cline{8-9} \cline{11-12} 
& Mean& Median& & Mean& Median& & Mean& Median& & Mean& Median\\
\hline
VLRG& 1.4\,$\pm$\,0.3& 1.3\,$\pm$\,0.2& & 1.9\,$\pm$\,0.4& 1.7\,$\pm$\,0.3& & 1.2\,$\pm$\,0.4& 1.2\,$\pm$\,0.3& & 0.8\,$\pm$\,0.2& 0.7\,$\pm$\,0.1\\
LRG& 1.6\,$\pm$\,0.3& 1.6\,$\pm$\,0.3& & 2.4\,$\pm$\,0.8& 2.3\,$\pm$\,0.4& &1.4\,$\pm$\,0.2& 1.3\,$\pm$\,0.1& & 1.6\,$\pm$\,0.2& 1.6\,$\pm$\,0.2\\
HRG& 1.5\,$\pm$\,0.3& 1.5\,$\pm$\,0.2& & 2.9\,$\pm$\,0.4& 2.7\,$\pm$\,0.4& & 1.4\,$\pm$\,0.2& 1.4\,$\pm$\,0.1& & 2.6\,$\pm$\,0.3& 2.4\,$\pm$\,0.2\\
VHRG& 1.7\,$\pm$\,0.2& 1.7\,$\pm$\,0.2& & 4.4\,$\pm$\,0.4& 4.6\,$\pm$\,0.4& & 1.9\,$\pm$\,0.2& 1.8\,$\pm$\,0.1& & 6.0\,$\pm$\,0.4& 6.2\,$\pm$\,0.3\\
\hline
Diffuse& 1.3\,$\pm$\,0.3& 1.2\,$\pm$\,0.2& & 2.2\,$\pm$\,0.4& 2.0\,$\pm$\,0.3& & 1.3\,$\pm$\,0.2& 1.2\,$\pm$\,0.2& & 1.5\,$\pm$\,0.3& 1.5\,$\pm$\,0.2\\
Moderate& 1.7\,$\pm$\,0.3& 1.6\,$\pm$\,0.2& & 2.8\,$\pm$\,0.4& 2.6\,$\pm$\,0.4& & 1.5\,$\pm$\,0.2& 1.5\,$\pm$\,0.1& & 2.4\,$\pm$\,0.3& 2.3\,$\pm$\,0.2\\
Compact& 1.7\,$\pm$\,0.2& 1.7\,$\pm$\,0.2& & 4.7\,$\pm$\,0.4& 4.8\,$\pm$\,0.4& & 1.9\,$\pm$\,0.1& 1.9\,$\pm$\,0.1& & 6.4\,$\pm$\,0.4& 6.3\,$\pm$\,0.4\\
\hline
Without SF& 1.4\,$\pm$\,0.3& 1.4\,$\pm$\,0.2& & 2.4\,$\pm$\,0.4& 2.3\,$\pm$\,0.4& & 1.3\,$\pm$\,0.2& 1.3\,$\pm$\,0.1 & & 2.4\,$\pm$\,0.3& 1.7\,$\pm$\,0.2\\
Possible SF& 1.5\,$\pm$\,0.3& 1.4\,$\pm$\,0.2& & 3.3\,$\pm$\,0.4& 3.0\,$\pm$\,0.4& & 1.7\,$\pm$\,0.2& 1.6\,$\pm$\,0.1& & 2.7\,$\pm$\,0.3& 2.2\,$\pm$\,0.3\\
With SF& 1.9\,$\pm$\,0.2& 1.9\,$\pm$\,0.2& & 4.2\,$\pm$\,0.4& 4.0\,$\pm$\,0.4& & 1.8\,$\pm$\,0.1& 1.9\,$\pm$\,0.1& & 5.2\,$\pm$\,0.3& 5.8\,$\pm$\,0.3\\
\hline
\end{tabular}
\label{table:categories}
\tablefoot{The classifications are described in Tables\,\ref{table:CO spectral parameters} and \ref{table:counts}.}
\end{table*}

In addition to the Milky Way, CO gas has been detected extending beyond the nominal $R_{25}$ radius\footnote{\tiny $R_{25}$ radius was taken from the NASA/IPAC Extragalactic Database (NED).} in a few galaxies, including NGC\,4414 ($\sim$\,13\,kpc), M\,33 ($\sim$\,9\,kpc), M\,51 ($R_{25}$\,$\sim$\,12\,kpc) and M\,83 ($R_{25}$\,$\sim$\,18\,kpc) (e.g., \citealt{Braine1993b,Braine2004,Koda2012,Koda2020,Druard2014}). A variable $R_{21}$ ratio has also been observed in the outer disks of some nearby galaxies. Specifically, the $R_{21}$ ratio is measured to be $\lesssim$\,0.5 at the optical edge of NGC\,4414 \citep{Braine2004}, and this ratio does not exhibit dependence on galactocentric distance \citep{Braine1993b}. 
Furthermore, the $R_{21}$ ratio in the outer disk of M\,33 is $\sim$\,0.8, showing no significant variation with galactocentric radius \citep{Druard2014}. The $R_{21}$ ratio demonstrates an increase towards the downstream side of the spiral arms, subsequently decreasing to 0.60 in the outer regions of M\,51 and M\,83 \citep{Koda2012,Koda2020}. 
Possibly as a consequence of the larger applied beam sizes, the $R_{21}$ ratios in the outer disks of nearby galaxies seem to be influenced by larger scale factors than our Galactic edge clouds. 

\subsection{Enhancement of $R_{21}$}
\label{Sec:Possible Causes}
The $R_{21}$ ratio can serve as a tracer of gas-compressed regions characterized by relatively high temperatures and densities (e.g., \citealt{Vaduvescu2007, Zhang2019}). 
Radiative transfer solutions of CO lines are detailed in Appendix\,\ref{Sect:Derivation}. The $R_{21}$  ratio ($R_{21}$\,$\sim$\,$T_{\rm mb}$(2--1)/$T_{\rm mb}$(1--0); see Sect.\,\ref{Sec:Relation} or Fig.\,\ref{fig:R-R}) converges towards unity under conditions of optically thick, warm, and dense gas assuming LTE. In scenarios of optically thin emission, the $R_{21}$ ratio could potentially reach up to values of four in dense and warm gas. 
Empirical evidence suggests that regions with high $R_{21}$ ratios are commonly associated with dense cloud and/or active star-forming regions. 

\subsubsection{Star-forming activity}
\label{Sect:Star-forming Activity}
Previous studies have probed high $R_{21}$ ratios as caused by massive star formation activity. 
GMCs that exhibit active star formation tend to have higher $R_{21}$ ratios, whereas those with quiescent star formation display lower ratios \citep{Sakamoto1994a,Nishimura2015,Yajima2021,Egusa2022}. 
Star formation activity has been identified within Digel Cloud 1 and 2 (see Sec.\,\ref{sect:Targets}). 
Two sources we observed, G131.016+1.524 and G131.157+1.390, located within Digel Cloud 1, are classified as moderate clouds (see Sect.\,\ref{sect:Overview}) and are associated with star-formation activity (see Table\,\ref{table:CO spectral parameters}).
Additionally, G137.759--0.983 and G137.775--1.067 situated within Digel Cloud 2, are compact clouds that exhibit a correlation with star-formation activity.
The molecular gas within the Digel Cloud is characterized by notably high $R_{21}$ ratios (0.69--1.43).
It is more likely for molecular gas to exhibit characteristics of being warm, dense, and optically thin when $R_{21}$\,>\,2.0 (for G117.576+3.950, G131.575+1.300, and G139.116--1.475). Nevertheless, two of the three sources are not associated with star-formation activity. The exeption is G117.576+3.950. 
The highest $R_{21}$ value, measured at 3.04, is observed in G139.116–1.475. 
This object is characterized by its compact structure and lack of star formation activity. 
The absence of notable star formation activity in G139.116--1.475 may hint at external heating by neighboring stars or it may be a sign of an extremely early phase in the star formation process. 

\begin{figure}[t]
\centering 
\includegraphics[width=0.49\textwidth]{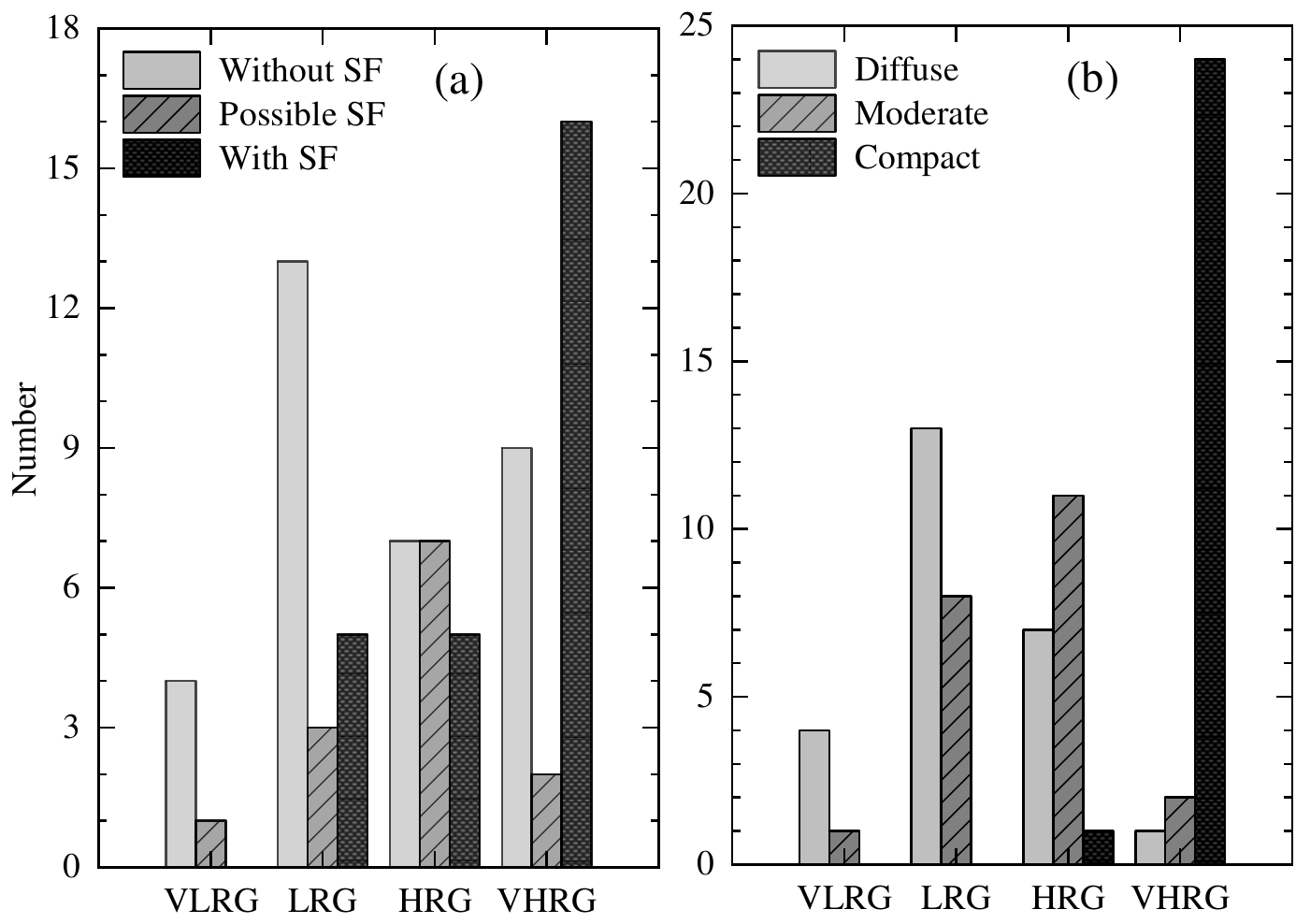} 
\caption{Histograms connecting $R_{21}$ (a) with the star formation properties and (b) with the compactness of our 72 targets. For the classification of VLRG, LRG, HRG, and VHRG clouds, see Sect.\,\ref{sect:CO-Line-Ratio}. For the determination of star-forming activity, see Sect.\,\ref{Sect:Star-forming Activity}, while Sect.\,\ref{Sect:Compact Clouds} refers to the overall structure of the clouds.}
\label{fig:R21-SF-Clumpy}
\end{figure}

\begin{figure*}[t]
\centering 
\includegraphics[width=1.0\textwidth]{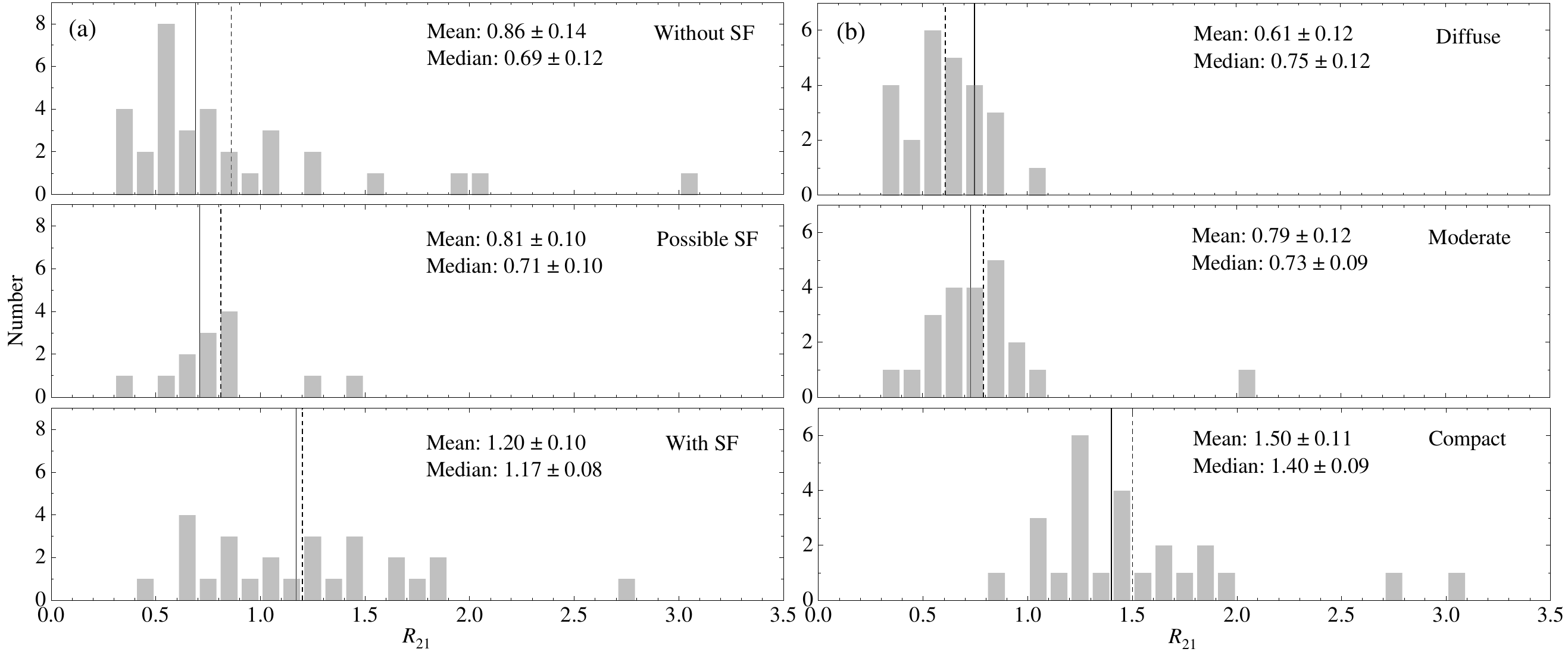} 
\caption{Histograms connecting $R_{21}$ (a) with the star formation properties and (b) with the compactness of our 72 targets. For the classification of VLRG, LRG, HRG, and VHRG clouds, see Sect.\,\ref{sect:CO-Line-Ratio}. For the determination of star-forming activity, see Sect.\,\ref{Sect:Star-forming Activity}, while Sect.\,\ref{Sect:Compact Clouds} refers to the overall structure of the clouds.
The dotted lines and the solid lines denote the mean and the median values, respectively.} 
\label{fig:structure}
\end{figure*}

We conduct a statistical analysis to explore the correlation between the $R_{21}$ ratio of 
Galactic edge clouds and their association with star-forming activity in Figs.\,\ref{fig:R21-SF-Clumpy} and \ref{fig:structure}. 
Based on the association between molecular clouds and star formation activity with the ${\it WISE}$-certified young stars (for details, see \citealt{Sun2015b}), 72 Galactic edge clouds were systematically classified into three distinct categories: with star formation, possible star formation, and without star formation (see Table\,\ref{table:CO spectral parameters}). 
This evaluation demonstrates that the HRG and VHRG molecular clouds are closely linked to star formation, as well as possible star formation (see Fig.\,\ref{fig:R21-SF-Clumpy}). 
Star formation activity is notably weak in the VLRG molecular clouds. 
Within Galactic edge clouds associated with star formation activity, the proportion of VHRG to the total molecular gas surpasses that in the clouds without star formation activity. 
It seems that Galactic edge clouds, which exhibit star formation activity, typically display higher $R_{21}$ ratios compared to those clouds devoid of such activity. 
In addition, only those relatively large and massive clouds may be detected at such distances. 
So a high $R_{21}$ is not unexpected. 

\begin{figure}[t]
\centering
\includegraphics[width=0.49\textwidth]{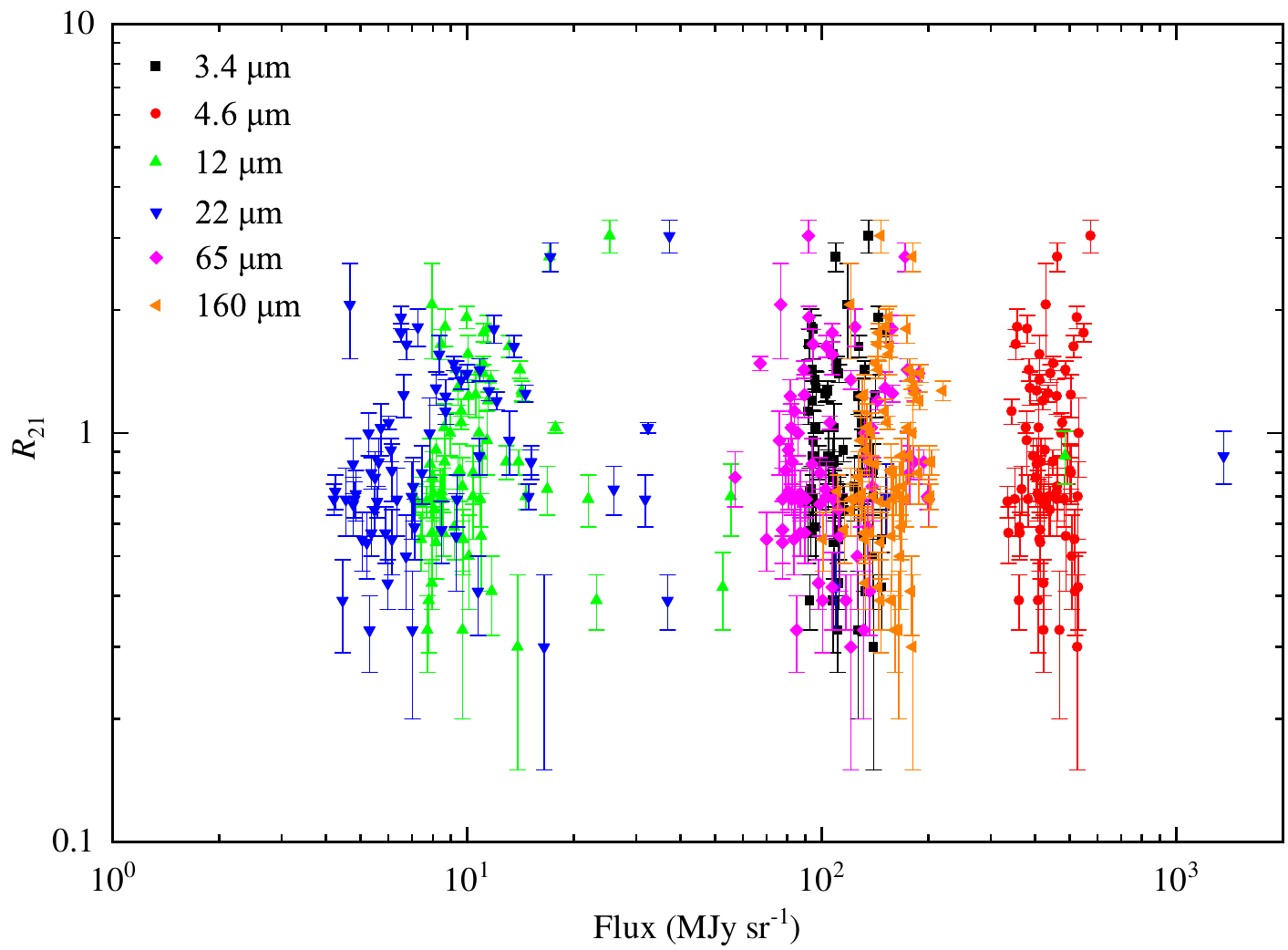}
\caption{No correlation between $R_{21}$ ratios and infrared flux.}
\label{fig:infrared flux}
\end{figure}

Previous observations suggest that the $R_{21}$ ratio exhibits a positive correlation with star formation rate indicators, including H$\alpha$ emission (pc scale) (e.g., \citealt{Egusa2022,Maeda2022}), the surface density of star-formation rate (kpc scale) (e.g., \citealt{Yajima2021,Leroy2022,Maeda2023,Jiang2024,Keenan2025}), and infrared emission (kpc scale) (e.g., \citealt{Braine1992,Koda2012,Koda2020,Yajima2021,denBrok2021}). 
This suggests that the relatively high $R_{21}$ ratio in the Galactic edge clouds may be considered on smaller scales with star-forming clouds, as well as on larger scales with star-forming galaxies.
However, we did not find any evidence to suggest that the $R_{21}$ ratios are correlated with infrared flux ({\it WISE} 3.4--22\,$\mu$m and {\it AKARI} 65--160\,$\mu$m, see Fig.\,\ref{fig:infrared flux}). 
In addition, $R_{21}$ depends on the gas kinetic temperature, while the infrared luminosity is connected to the dust temperature. 
This suggests that there is not a close correlation between the kinetic temperature and the dust temperature, due to low gas densities. 
The $R_{21}$ ratios are not significantly linked to metallicity gradients with Galactocentric distance in the outer Galaxy (see Sect.\,\ref{Sec:Variation} or Fig.\,\ref{fig:Rg-R21}). 
The variable $R_{21}$ values appear to correspond to molecular clouds at different evolutionary stages of the star-forming process at the edge of the Galaxy. 

\subsubsection{Clumpy structure}
\label{Sect:Compact Clouds}
In low-metallicity environments, diffuse areas within molecular clouds may lack sufficient dust to protect CO molecules from photodissociation. 
Observations have shown high $R_{21}$ ratio gas with compact components, as well as low ratio gas with diffuse components (e.g., \citealt{Sakamoto1994a,Usuda1999,Hasegawa1997a,Sorai2001,Nishimura2015}). 
The $R_{21}$ ratios of our Galactic edge clouds were statistically examined
by categorizing the compactness levels of molecular clouds (diffuse, moderate, and compact; see Sect.\,\ref{sect:Overview}) in Figs.\,\ref{fig:R21-SF-Clumpy} and \ref{fig:structure}. 
The compact clouds tend to exhibit higher $R_{21}$ ratios than the diffuse and moderate clouds. 
Previous observations indicate that the fraction of CO-dark H$_2$ to total H$_2$ increases with Galactocentric distance, ranging from $\sim$20\% at 4\,kpc to $\sim$80\% at 10\,kpc \citep{Pineda2013}. In the solar neighborhood, approximately half of the molecular gas is CO-dark (e.g., \citealt{Paradis2012,Pineda2013,Chen2015}). This contribution should be even more significant in the low metallicity regions of the Galactic edge clouds \citep{Pineda2013,Langer2014,Luo2024}. As a result, only regions with particularly high column density are seen in CO emissions, thereby enhancing the $R_{21}$ ratio.

\subsubsection{Linewidths, compactness, and brightness temperatures} 
\label{Sec:Linewidth}
Regardless of CO\,(1--0) or (2--1), both the mean and median linewidths demonstrate a progressive increase, ranging from VLRG to VHRG (see Table\,\ref{table:categories}). 
Similarly, average CO linewidth values are higher in compact clouds compared to those in diffuse clouds. 
Furthermore, the linewidths are broader in the Galactic edge clouds that exhibit star formation compared to those devoid of such activity (see Table\,\ref{table:categories}). 
Nevertheless, the CO linewidths of our Galactic edge clouds are 
systematically narrow with a mean value of $\sim$\,1.6\,km\,s$^{-1}$ (see Sect.\,\ref{sect:Overview} and Table\,\ref{table:CO spectral parameters}). 
This is similar to observational results from CO in infrared dark clouds (e.g., \citealt{Li2016}) but much smaller than typical values
($\sim$\,few\,km\,s$^{-1}$) observed with multiple molecular species in massive star-forming regions of our Galaxy and the LMC (e.g., \citealt{Tang2013,Tang2014,Tang2017a,Tang2017b,Tang2018a,Tang2018b,Tang2021,Giannetti2014,Giannetti2017,Gong2023,Green2024,Zhao2024}). 
The brightness temperature of CO lines exhibits an increase from diffuse to compact clouds within the Galactic edge clouds. 
This trend is also observed in the transition from low- to high-ratio gas of $R_{21}$ (see Table\,\ref{table:categories}). 
It may imply that the majority of high $R_{21}$ clouds in the extreme outer Galaxy are more likely to form stars due to their intrinsic density, rather than due to external pressure. 

\subsubsection{Galactic arm} 
\label{Sec:arm}
The radial variation of $R_{21}$ may be caused by the mixing ratio variation of molecular gas with different $R_{\rm g}$ \citep{Sakamoto1997}. 
The $R_{21}$ ratio increases from <\,0.7 in the interarm regions to >\,0.7 in the spiral arms (e.g., \citealt{Koda2012,Koda2020,Maeda2022}), indicating a substantial 
fraction of the high $R_{21}$ gas in the spiral arms and a larger fraction of the low $R_{21}$ gas in the interarm regions. 
The Galactic edge clouds we observed are situated on a new arm (see Sect.\,\ref{sect:Galactic-Edge-Clouds}), and the enhancements of the $R_{21}$ ratio on scales of 2.3--3.7\,pc are attributed to this region. 
Considering these factors, it appears that the $R_{21}$ radial trend in the Galaxy is likely due to arm-like structures and/or mixing ratios of multiple $R_{21}$ gases. 
However, variation of $R_{21}$ in the Galactic edge clouds is likely caused by multi-scale and multi-physical processes, and a simple explanation is not entirely convincing. 
Therefore, further studies involving more samples and diverse molecular probes are anticipated for the investigation of Galactic edge clouds.

\section{Summary}
\label{sect:summary}
We conducted observations of CO\,($J$=2--1) spectral lines towards 72 molecular clouds in the Galactic edge at Galactocentric distances of $R_{\rm g}$\,=\,14--22\,kpc, utilizing the IRAM\,30\,m telescope. 
By integrating the CO\,($J$=1--0) data obtained from the MWISP project, we investigated the variations of the $R_{21}$ 
ratios across these Galactic edge clouds. The main results are the following: 
\begin{enumerate}
\item
CO\,(2--1) has been detected in all observed 72 sources. 
With a resolution of 52$^{\prime\prime}$ and accounting for beam size effects, the $R_{21}$ ratio values derived from CO\,(2--1) to (1--0) integrated intensity ratios span from 0.30 to 3.04 with a mean of 0.97\,$\pm$\,0.12 and a median of 0.81\,$\pm$\,0.10 in 72 Galactic edge clouds. 
The proportions of VLRG, LRG, HRG, and VHRG (see Sect.\,\ref{sect:CO-Line-Ratio}) are found to be 6.9\%, 29.2\%, 26.4\%, and 37.5\%, respectively. 
This indicates a significant presence of molecular gas with a high proportion and elevated excitation levels within the Galactic edge clouds. 

\item
The $R_{21}$ ratio in our Galaxy shows a gradient of initial radial decline followed by a high dispersion in Galactocentric distance, akin to that observed in NGC\,2903. The scattering $R_{21}$ ratio does not display a systematic variation within the range of $R_{\rm g}$\,=\,14--22\,kpc at the periphery of the Milky Way. 

\item
The high proportion of HRG and VHRG in our edge clouds is correlated with compact clouds and regions exhibiting star-forming activity. 
This implies that the high $R_{21}$ ratios could be attributed to dense gas concentrations and recent episodes of star formation. 
\end{enumerate}

The Galactic edge clouds present an ideal location for investigating molecular clouds and  star formation within low metallicity
environments. Future research will focus on the comprehensive investigation of the physical and chemical characteristics of the 
Galactic edge clouds. 

\begin{acknowledgements}
The authors thank the referee for helpful comments. 
We thank the staff of the IRAM telescope for their assistance in observations.
This work acknowledges the support of the National Key R\&D Program of China under grant Nos.\,2023YFA1608002 and 2022YFA1603103, the Tianshan Talent Training Program of Xinjiang Uygur Autonomous Region under grant No.\,2022TSYCLJ0005, the Chinese Academy of Sciences (CAS) “Light of West China” Program under grant No.\,xbzg-zdsys-202212, and the Natural Science Foundation of Xinjiang 
Uygur Autonomous Region under grant No.\,2022D01E06. It was also partially supported by 
the Regional Collaborative Innovation Project of XinJiang Uyghur Autonomous Region under grant No.\,2022E01050,
the National Natural Science Foundation of China under grant Nos.\,12173075, 12373029, and 12463006, the Xinjiang Key Laboratory of Radio Astrophysics 
under grant No.\,2023D04033, the Natural Science Foundation of Xinjiang Uygur Autonomous Region under grant No.\,2022D01A359, 
and the Youth Innovation Promotion Association CAS.  C.\,Henkel acknowledges support by the Chinese Academy of Sciences President's 
International Fellowship Initiative under grant No.\,2025PVA0048. X.\,P.\,Chen, T.\,Liu, K.\,Wang, and J.\,W.\,Wu acknowledge support by 
the Tianchi Talent Program of Xinjiang Uygur Autonomous Region. MWISP is sponsored by the National Key R\&D Program of China with grant 2023YFA1608000
and the CAS Key Research Program of Frontier Sciences with grant QYZDJ-SSW-SLH047. This research has used NASA's Astrophysical Data System (ADS).
\end{acknowledgements}

\bibliography{co_ratio} 

\begin{appendix}
\onecolumn
\section{Typical structure of the Galactic edge clouds}
\label{Sect:Typical molecular structure}
\begin{figure*}[h]
\centering
\includegraphics[width=0.99\textwidth]{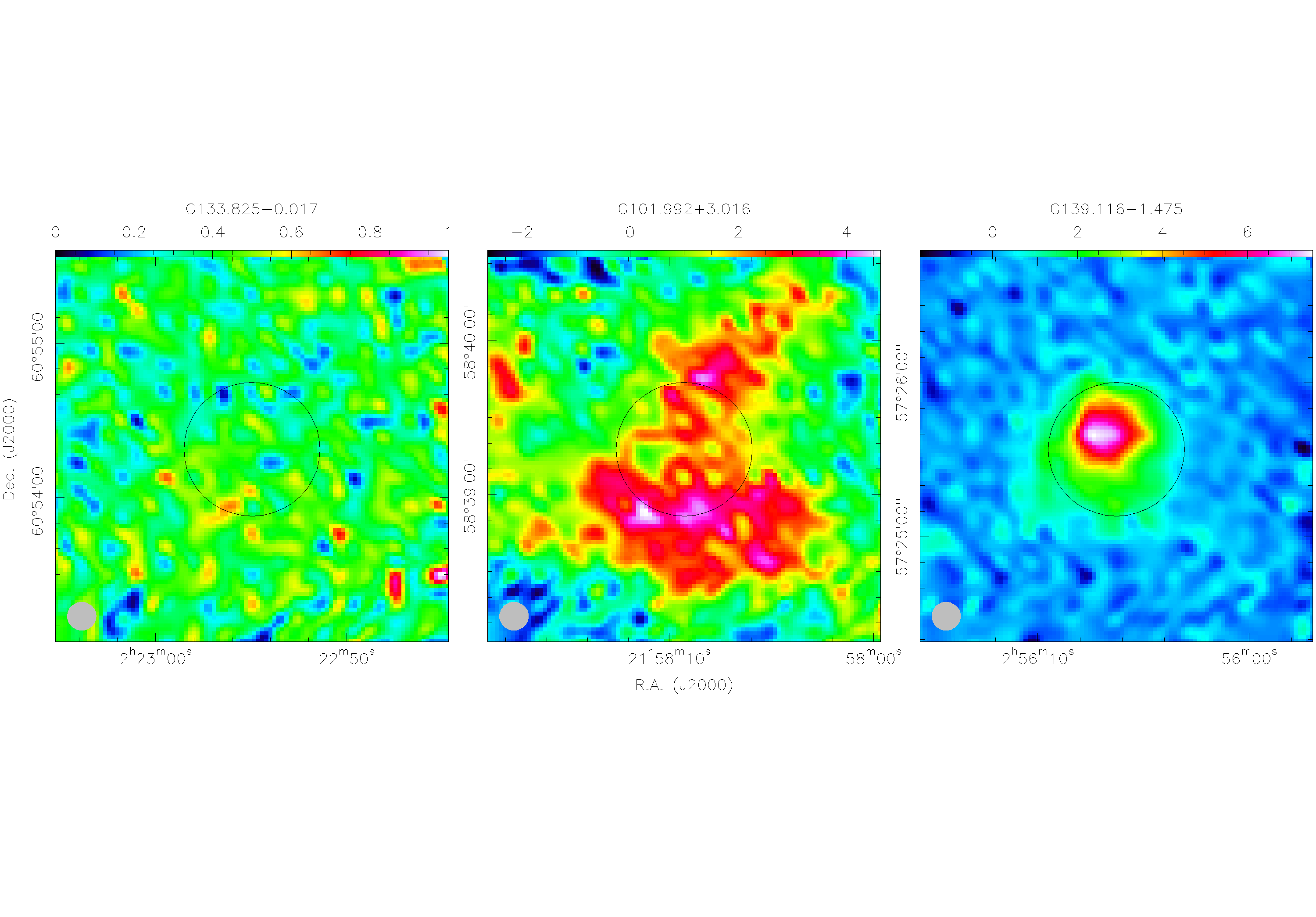}
\caption{Typical velocity-integrated intensity maps of CO\,(2--1) for diffuse (\emph{left}), moderate (\emph{middle}), and compact structures (\emph{right}) of Galactic edge clouds ($T_{\rm mb}$ scale; color bar in units of K\,km\,s$^{-1}$). The black circles in the center of each map represent the CO\,(1--0) emission peaks identified from the Delingha 13.7\,m and their size corresponds to the telescope beam size of $\sim$\,52$^{\prime\prime}$. The grey filled circles in the lower left corner show the beam size of CO\,(2--1) observed with the IRAM\,30\,m telescope. 
}
\label{fig:Typical molecular structure}
\end{figure*}

\twocolumn
\section{Derivation of the CO\,$J$=2--1/1--0 line ratio}
\label{Sect:Derivation}
In the context of LTE, the solution of the radiative transfer equation is formulated in a manner that involves the radiation brightness temperature of the target \citep{Mangum2015} 
\begin{equation}
T_{\rm mb}=(1-e^{-\tau_\nu})J_{\nu}(T_{\rm ex}), \label{eq:T_mb}
\end{equation}
where $\tau_\nu$ represents the optical depth of the transition. 
A radiation temperature, often specified in the millimeter and submillimeter range, is defined as follows
\begin{equation}
J_{\nu}(T_{\rm ex})=\frac{h\nu}{k} \frac{1}{e^{h\nu/kT_{\rm ex}}-1}, \label{eq:J_Tex}
\end{equation}
where $T_{\rm ex}$ denotes the excitation temperature. $h$ and $k$ are the Planck and Boltzmann constants. 
For linear molecules, when the centrifugal stretching constant is neglected, transition frequencies are multiples of $\nu$\,$\sim$\,$2BJ$. 
$B$ is the molecular rotation constant and $J$ represents the rotational quantum number of the upper level of the transition. 
The cosmic microwave background temperature is small with respect to $T_{\rm ex}$ and can be neglected for the sake of simplification. 
The brightness temperature ratio for two consecutive transitions, specifically CO\,(2--1) and CO\,(1--0), is denoted by 
\begin{equation}
\frac{T_{\rm mb}(2-1)}{T_{\rm mb}(1-0)} = 2 \frac{1-e^{-\tau_{21}}}{1-e^{-\tau_{10}}} \frac{e^{\frac{h\nu_{10}}{kT_{\rm ex,10}}}-1}{e^{\frac{h\nu_{21}}{kT_{\rm ex,21}}}-1}, \label{eq:thick}
\end{equation}
where all indices 21 denote physical quantities associated with CO\,(2--1), whereas indices 10 signify physical quantities linked to CO\,(1--0). 
For example, $\tau_{21}$ and $\tau_{10}$ represent the optical depths of the CO\,(2--1) and CO\,(1--0) transitions. 
According to the assumption of LTE, the excitation temperature $T_{\rm ex,21}$ equals $T_{\rm ex,10}$, which in turn is equal to the kinetic temperature $T_{\rm kin}$. In optically thick ($\tau_{\nu}$\,$\gg$\,1), warm ($kT_{\rm kin}$\,$\gg$\,$h\nu$), and dense ($n_{\rm H_2}$\,>\,10$^{3}$\,cm$^{-3}$) gas, the $T_{\rm mb}$(2--1)/$T_{\rm mb}$(1--0) ratio approaches unity. 

Under the Rayleigh-Jeans approximation ($h\nu/kT_{\rm ex}$\,$\ll$\,1), the optical depth is simply represented by
\begin{equation}
\tau_{\nu} \propto \frac{J}{2J+1} \nu n, \label{eq:tau}
\end{equation}
where $n$ denotes the number of molecules in a specific energy level.
In a system that is in thermal equilibrium, the relative level populations adhere to the Boltzmann distribution
\begin{equation}
\frac{n_u}{n_l} = \frac{g_u}{g_l}e^{-\frac{h\nu}{kT}}, \label{eq:boltz}
\end{equation}
where $g_l$ and $g_u$ represent the degeneracy of the lower and upper energy levels of a transition, specifcally (2$J$+1) for CO. The energy separation for these two transitions are $hv_{10}/k_B$\,=\,5.5\,K and $hv_{21}/k_B$\,=\,11.04\,K, respectively. 
In the case of optically thin emission ($\tau_{\nu}$\,$\ll$\,1), Eq.\,\ref{eq:thick} can be reduced to
\begin{equation}
\begin{split}
\frac{T_{\rm mb}(2-1)}{T_{\rm mb}(1-0)} &\approx \frac{\tau_{21}}{\tau_{10}} \approx 4 e^{-\frac{11.04}{T_{\rm ex,21}}} , \label{eq:thin}
\end{split}
\end{equation}
implying that the $T_{\rm mb}$(2--1)/$T_{\rm mb}$(1--0) ratio can potentially attain a value as high as four in dense, warm, and optically thin gas.
The $T_{\rm mb}$(2--1)/$T_{\rm mb}$(1--0) ratio exhibits an exponential dependence on the excitation temperature at low temperatures. 
This ratio may approach unity under optically thick conditions, while it may reach values up to four in optically thin conditions. These conclusions are also valid for the $R_{21}$ ratio of velocity integrated intensities, if both transitions have the same line widths. 

\end{appendix}
\end{document}